\documentclass[12pt]{iopart}   

\usepackage{iopams}
\usepackage[english]{babel}
\usepackage{graphicx}
\usepackage{dcolumn}
\usepackage{bm}
\usepackage{verbatim}
\usepackage{amsfonts}
\usepackage{amssymb}
\usepackage{color}
\usepackage{enumerate}
\newcommand{\be}[1]{\begin{equation}\label{#1}}
\newcommand{\ee}{\end{equation}}
\newcommand{\bea}[1]{\begin{eqnarray}\label{#1}}
\newcommand{\eea}{\end{eqnarray}}
\newcommand{\no}{\nonumber \\}
\newcommand{\Fig}[1]{Fig.(\ref{#1})}
\newcommand{\Eq}[1]{Eq.(\ref{#1})}
\newcommand{\Sec}[1]{Section~\ref{#1}}
\newcommand{\App}[1]{Appendix~\ref{#1}}
\newcommand{\bsub}{\begin{subequations}}
\newcommand{\esub}{\end{subequations}}

\def\np{{n_p}}
\def\np0{{n_{p0}}}

\def\Tr{\trm{Tr}}
\def\Id{\hat{\mathbf{1}}}
\def\ep{\epsilon}

\def\Xakinpm{X^{\pm}_{a^{in}_{k}}}

\def\Xakoutpm{X^{\pm}_{a^{out}_{k}}}
\def\Xakoutp{X^{+}_{a^{out}_{k}}}
\def\Xakoutm{X^{-}_{a^{out}_{k}}}
\def\Vakinpm{V^{\pm}_{a^{in}_{k}}}
\def\Vakpm{V^{\pm}_{a^{out}_{k}}}
\def\Vakp{V^{+}_{a^{out}_{k}}}
\def\Vakm{V^{-}_{a^{out}_{k}}}
\def\Xbkpm{X^{\pm}_{b_{k}}}

\def\Xbkmpm{ X^{\pm}_{b_{k-1}} }

\def\Vbkpm{V^{\pm}_{b_{k}}}
\def\Vbkp{V^{+}_{b_{k}}}
\def\Vbkm{V^{-}_{b_{k}}}
\def\Vbkmpm{V^{\pm}_{b_{k-1}}}

\def\Vbzpm{V^{\pm}_{b_{0}}}

\def\Vbzkm{V^{-}_{b_{0}}}
\def\Rk{R_{k}}
\def\Rkm1{R_{k-1}}
\def\Nak{N_{a_k}}
\def\Nbk{N_{b_k}}
\def\Nbkm{N_{b_{k-1}}}
\def\Nb0{N_{b_0}}
\def\rak{r_{a_k}}
\def\rbk{r_{b_k}}
\def\erakmp{e^{\mp 2\,r_{a_k}}}
\def\erbkmp{e^{\mp 2\,r_{b_k}}}
\def\Id{\mathbf{1}}
\def\bsig{\mathbf{\sigma}}

\usepackage{color}

\def\trm#1{\textrm{#1}}
\def\tit#1{\textit{#1}}

\newcommand{\ket}[1]{|#1\rangle}
\newcommand{\bra}[1]{\langle #1|}

\def\SBekHawk{S_{\trm{\tiny Bekenstein-Hawking}}}
\def\quarter{\frac{1}{4}}
\def\Tr{\trm{Tr}}
\newcommand{\appropto}{\mathrel{\vcenter{
  \offinterlineskip\halign{\hfil$##$\cr
    \propto\cr\noalign{\kern2pt}\sim\cr\noalign{\kern-2pt}}}}}
\newcommand{\defn}{\mathrel{\vcenter{
  \offinterlineskip\halign{\hfil$##$\cr
    {\scriptsize \trm{def}}\cr\noalign{\kern2pt}=\cr\noalign{\kern-2pt}}}}}
\newcommand{\mylimit}[2]{\mathrel{\vcenter{
  \offinterlineskip\halign{\hfil$##$\cr
    {\scriptstyle {#1}}\cr\noalign{\kern2pt}{#2\;\;}\cr\noalign{\kern-2pt}}}}}    

\begin{document}


\title[Quantum Optical Inspired Models for Unitary Black Hole Evaporation]{Quantum Optical Inspired Models for Unitary Black Hole Evaporation}
%

\author{Paul M. Alsing }
\address{Florida Atlantic University, 777 Glades Rd, Boca Raton, FL, 33431}
\ead{palsing@fau.edu,alsingpm@gmail.com}
\date{\today}

\begin{abstract}
\noindent In this work, we describe  optically inspired models for unitary black hole (BH) evaporation. 
The goal of these models are 
(i) to be operationally  simple, 
 (ii)~approximately preserve the thermal nature of the emitted Hawking Radiation (HR), and 
 (iii) attempt to reproduce the Page Curve
that purports that information flows forth from the BH 
when it has evaporated to approximately half its initial mass.
We concentrate on modeling the BH as a single mode squeezed state successively interacting, by means of beam splitters and squeezers, with vacuum modes near the horizon, giving rise to entangled pairs representing the external Hawking radiation and its partner particle inside the horizon. Since all states and operations are Gaussian throughout, we use a symplectic formalism to track the evolution of the composite system through the evolving means and variances of their quadrature operators. This allows us to easily compute correlations and entanglement  between the BH and the HR, as well as calculate correlations between the BH at early and late times.
\end{abstract}

\section{Introduction}\label{sec:intro}
Hawking's celebrated result that black holes (BH) evaporate \cite{Hawking:1975} has now entered its sixth decade of continued active investigation \cite{Page:2005}. Central to this result is the BH information problem, which in essence questions the unitarity of BH evaporation if an incoming pure state evolves to an outgoing Hawking radiation thermal (mixed) state, when the BH has completely evaporated. 
The thermal nature of the Hawking radiation arises from the highly entangled nature of the vacuum, which in the case of the maximally extended Schwarzschild spacetime is envisioned to be the thermofield double (TFD) state, 
$\ket{TFD} = Z^{-1} \sum_{n=0}^\infty e^{-\beta\,E_n}\,\ket{n}_L\otimes\ket{n}_R$ \cite{Almheiri_Hartman:2020,Almheiri_Hartman:2021}, where 
$Z = \sum_{n=0}^\infty e^{-\beta\,E_n}$  is the partition function, and $L$ and $R$ are the causally disconnected left and right Rindler wedges of the associated Penrose diagram. In the external region of the BH in ``our universe," (the R-wedge) the reduced quantum state of the of the Hawking radiation $\rho_{R}$ is given by tracing out over the causally disconnected region $L$ wedge, since no signal from the latter can enter the $R$ wedge. This yields 
$\rho_{R} = Z^{-1} \sum_{n=0}^\infty e^{-\beta\,E_n}\,\ket{n}_R\bra{n}$ which is a thermal state with Boltzmann probabilities $p_n = e^{-\beta\,E_n}/Z$, with accompanying von Neumann \tit{fine-grained}  entropy 
$S_{vN} = -\Tr[\rho_{R}\log \rho_{R}] =  -\Tr[\sum_n p_n\log p_n]$. 
%

However, if the BH begins in a pure state, and the emitted Hawking radiation is thermal \cite{Hawking:1975}, then these conditions are at odds with the decreasing  coarse-grained 
Bekenstein-Hawking thermodynamic entropy  given by $\SBekHawk = \quarter A_{BH}$ (in units of the Planck length ($L^2_P \defn \hbar\,G/c^3$) squared). 
This is because the thermodynamic Bekenstein-Hawking  entropy $\SBekHawk$ is a \tit{coarse-grained} entropy, which is a maximum of the fine-grained von Neumann entropies over all density matrices describing the system with a limited set of observables involved \cite{Almheiri_Hartman:2021}. This implies that we must always have $S_{vN} \le \SBekHawk$.
However, at some midpoint in the BH evaporation process, called the Page Time \cite{Page:2005,Page:1993a,Page:1993b}, the area of the BH has shrunk to where it has emitted roughly half its mass to the external Hawking radiation and
 $S_{vN} \approx \SBekHawk$, after which $S_{vN}$ becomes inconsistently less than $\SBekHawk$.
 Thus, the generalized entropy $S_{gen}$ consisting of the $\SBekHawk$ of the BH plus the outside entropy of the external Hawking radiation (and any other quantum fields) must ``turn over" at the Page time, and once again decrease, eventually to zero, as the the BH evaporates and the Hawking radiation returns to a pure state. It is at this midpoint, where the decreasing Hilbert space dimension of the BH is roughly equal to the Hilbert space dimension of the external radiation, that information begins to flow out of the BH, i.e. where the entanglement entropy between the inside and outside of the BH is maximal. 
 This is given by the celebrated Page Curve, as shown in \Fig{Page:S_I_vs_lnm}, and discussed in the next section.  It has been reported in Harlow's lectures on BH and Quantum Information \cite{Harlow:2016} that Andy Strominger has argued that being able to produce the Page Curve in some particular theory is what it means to have solved the BH information problem (see also p6-68 of Ydri \cite{Ydri:2025}). 
 
 Within the last 15-20 years, there has been much new insight gained into the computation of the generalized entropy $S_{gen}$ of the combined BH/HR system. These are lucidly discussed in the review article by Almheir \tit{et al}. \cite{Almheiri_Hartman:2020}, and in textbook detail by Ydri \cite{Ydri:2025}, and quite illuminating and informatively explained for the non-expert in the popular  book on BHs by Cox and Forshaw \cite{Cox_Forshaw:2022} (see also p1-9 of \cite{Alsing_CQG:2025} for a brief summary of these ideas). These involve new concepts including the Ryu-Takayangi formula for the area of quantum extremal surfaces-generalizing Bekenstein-Hawking formula \cite{Ryu_Takayanagi:2006,HRT:2007}, 
 the ``ER=EPR" proposal that wormholes are intimately related to the concept of quantum entanglement \cite{Maldacena_Susskind:2013}, and Entanglement Wedges inside the horizon supply additional contributions to the  the fine-grained entropy of the Hawking radiation \cite{Almheiri:2020,Penington:2020}.
In brief, these new results support a mechanism for microscopic wormholes arising from the quantum entanglement of the interior and exterior of the BH to allow  for evolving regions \tit{behind} the BH horizon to be counted in the computation of the fine-grained entropy of the external Hawking radiation, thus reducing the entanglement entropy, as the system returns to a pure state. Such concepts arise from a Euclidean path integral formulation of the computation of $S_{gen}$ \cite{Almheiri_Hartman:2020,Penington:2022} which then allow it to be computed in a semi-classical approximation (i.e. a classical GR background geometry plus quantum fields).

Within the last 10 years, an alternative quantum-optical-based approach to BH evaporation has been explored by Nation and Blencowe \cite{Nation_Blencowe:2010}, Adami and collaborators \cite{Adami_VerSteeg:2014,Bradler_Adami:2016}, and this author \cite{Alsing_CQG:2025,Alsing_CQG:2015,Alsing_Fanto_CQG:2016}.
The essence of the idea is that (i) the thermofield double state $\ket{TFD}$ discussed above is precisely the two mode squeezed vacuum state produced in the un-depleted pump laser approximation, discussed in any textbook on quantum optics \cite{Scully_Zubairy:1997,Agarwal:2013,Gerry_Knight:2023,Rice:2025}. The central idea is to model the BH as an effective ``pump laser," with the down converted signal and idler (photons/bosons) acting as the external Hawking radiation (HR) and its internal, behind the horizon, Hawking partner particle (HRPP), respectively. In the ``un-depleted pump" approximation, the BH does not evaporate.
Adami and collaborators pursued this analytically tractable model to explore both the classical information channel capacity of the BH \cite{Adami_VerSteeg:2014}, as well as obtaining the Page curve from a dynamical one-shot decoupling model \cite{Bradler_Adami:2016} (in which the signal/idler - HR/HRPP move away from the BH once they are created - much like a real laboratory quantum optical process of entangled bi-photon generation).
Nation and Blencowe \cite{Nation_Blencowe:2010} considered the case of a trilinear Hamiltonian, whereby the pump/BH is also quantized, thus allowing it to deplete  during signal/idler generation, thus modeling  
BH evaporation. There is no closed form expansion of the unitary operator formed from the trilinear Hamiltonian, so the authors  examined the BH evaporation/Hawking radiation perturbatively  in the Heisenberg picture, also generating a Page curve.
The prior works of this author \cite{Alsing_CQG:2015,Alsing_Fanto_CQG:2016}, inspired by the previous researchers (and working in the Schr\"{o}dinger picture) were also able to explore the short and longer term evolution process of BH evaporation, with the BH beginning in a quantum state of a large, but finite (boson) occupation (mean) number, again producing a Page curve. Modifications of the entanglement between the BH and the Hawking partner particles in this tripartite system, was also explored. The author's most recent work \cite{Alsing_CQG:2025} incorporated a cascading ``waterfall" process by which it was proposed that the internal infalling idler/HRPP itself generates subsequent signal/idler pairs, ad-infinitum, until the BH complete evaporates, and the system returns to a pure state, again producing a Page curve.
In a sense, this latter model  analogously mimics, to an degree, the operational functionality of the ER=EPR wormholes proposed by modern the approach to BH evaporation and Page curve generation discussed above, by also effectively ``moving" Hawking partner particles from inside the BH horizon, to outside. 

The key point of these trilinear Hamiltonian models is that the effective Hilbert space dimension of the trilinear system initially grows (from zero, in a composite pure state) as the pump/BH depletes, producing maximally entangled signal-idler/HR pairs. 
The BH begins in a pure quantum state with a  huge, but finite occupation number (or coherent state) while the HR/HRPP are taken to start in the vacuum, with zero entanglement entropy. 
As the BH evaporates, and its mean occupation number falls, and HR/HRPP occupation number grows, the Hilbert space of the respective subsystems becomes nearly equal, corresponding to the Page time, and the BH and HR/HRPP are at peak entanglement. 
Subsequently, as the BH occupation number marches towards zero as it continues to evaporate, and the occupation number of the HR/HRPP pairs continues to grow, the Hilbert space dimension of both systems begin to decrease once again, and both systems end up in pure states once again, with zero entanglement entropy.
Thus, the evolution of the trilinear Hamiltonian system mimics the Page curve, and the two-mode squeezed state inherently encodes the signal-idler/HR-HRPP entanglement which is naturally nearly thermal for early times. Therefore, these quantum optical inspired models of BH evaporation appear to contain the essential features one would want in a more formal, general theory. The goal of this present work is to explore another reasonable quantum optical model for BH evaporation , this time modeling the BH large occupation number single-mode squeezed state interacting at the horizon with vacuum modes through a combined beam splitter (BS) and subsequent re-squeezing process, in order to generate a Page curve.
As stated in the abstract, the objective of these models are threefold:  (i) to be operationally  simple, 
 (ii) approximately preserve the thermal nature of the emitted Hawking Radiation (HR), and (iii) attempt to reproduce the Page Curve.

The outline of this paper is as follows.
In \Sec{Page} we review Page's original formulation of his Page curve.
In \Sec{sec:model} we describe our beam splitter (BS) plus squeezing (SQ) model which we abbreviate as the  BSSQ model. Since the optical process of beam splitting and squeezing are all Gaussian, the initial Gaussian states remain thermal throughout the evolution process. We show how the system can be exactly evolved (event-wise)  in this case by a symplectic formalism that systematically tracks the evolution of the mean and variances of the Heisenberg quadratures (formally position and momentum operators). This formalism also allows for one to easily compute entanglement of subsystems, and as importantly, correlations between early and late times.
In \Sec{sec:thermal:states} we first examine the role of the BS alone (no squeezing), and its effect on the the outgoing HR. The free parameters in this model are the set of successive BS reflectivities 
$0\le \{R_k\}\le 1$ at iteration $k$, governing how much HR escapes from the BH to the outside.
In \Sec{sec:BS:and:SQ} we turn to our full BSSQ model. We examine two reasonable cases for the
reflectivities: (i) constant, and (ii) proportional to the BH occupation number/mass. We examine their effect on the subsequent evolution of the combined system, and the generation of Page curves.
We also compute correlations between the BH at early and late times by means of a measure of purity of the quantum density matrix, as well as entanglement by means of the Log Negativity.
Lastly, in \Sec{sec:conclusion:future:dirs} we present our conclusions and prospects for future directions of research.
 Finally, we include in the 
 \App{app} 
 we include the details of the formulas used to compute the correlations between the BH and HR quadrature operators.

\section{On the Page Information in the BH radiation}\label{Page}
The generally accepted conventional wisdom for when information leaks out of the BH stems from the seminal 1993 work of Page  in which he calculated (i) the average information in a subsystem \cite{Page:1993a} and (ii) then applied this to the information in the BH radiation \cite{Page:1993b}. The main result of this work is the \textit{Page time} which is roughly half the evaporation time of the BH  when the information in the outgoing Hawking radiation becomes appreciable. Here we briefly review the main results of Page's  1993 works, which conjectures that the information in the BH will begin to emerge when it has evaporated to approximately half its original mass.

\subsection{Review of Page's results}
In the first paper, \textit{The average entropy of a subsystem} \cite{Page:1993a}, Page considers a random pure state of fixed dimension $N= m n$ selected from the Haar measure, which is proportional to the standard geometric hypersurface volume on the unit sphere
$S^{2 m n}-1$ which those unit vectors give when the $m n$-complex-dimensional Hilbert
space is viewed as the $2 m n$-real-dimensional Euclidean space. The integer divisors $m,n$ of $N$ are considered as the dimensions of subsystems $A$ and $B$ respectively, taking without loss of generality,
$m < n$. The goal is to compute the average entropy
$\langle S_A \rangle = S_{m,n}$ (with respect to the Haar measure) and the information $I_{m,n} = (S_A)_{max} - \langle S_A \rangle = \ln m - S_{m,n}$. The end result of this work is the following \cite{Page:1993a}:
$S_{m,n} = \sum_{k=n+1}^{m n} 1/k - (m-1)/(2n)\approx \ln m - m/(2n)$ for $1\ll m \le n$.
The conclusion drawn from this work is that for a typical pure state of the composite system, (i) very little of the information, roughly $m/(2n)$ units, are in the correlations within the smaller subsystem A itself, (ii) roughly $\ln n- \ln m + m/(2n)$ units are in the correlations within the larger subsystem B itself, and (iii) the remaining roughly $2 \ln m - m/n$ units of information are in the correlations between the larger
and smaller subsystems.

In the second paper \textit{Information in black hole radiation} \cite{Page:1993b} subsystem $A$
of dimension $m$ is taken to be the Hawking radiation of the BH, system $B$ of dimension $n$. The  subsystems $A$ and $B$ are assumed to be correlated via the composite random pure state $\ket{\psi}_{A B}$, such that $\rho_A = Tr_B[\ket{\psi}_{AB}\bra{\psi}]$, and
$\rho_B = Tr_A[\ket{\psi}_{AB}\bra{\psi}]$.
 To model this, Page considers a large integer $N = m n = 291,600$, which has 105 integer factors $m$.
 For $m\le n$ he computes $S_{m,n} = \sum_{k=n+1}^{m n} 1/k - (m-1)/(2n)$, while for $m > n$ he uses
 $S_{m,n} = \sum_{k=m+1}^{m n} 1/k - (n-1)/(2 m)$, while in both regimes $I_{m,n} = \ln m - S_{m,n}$. This is plotted in \Fig{Page:S_I_vs_lnm} below.
\begin{figure}[ht]
\begin{center}
\includegraphics[width=4.0in,height=1.95in]{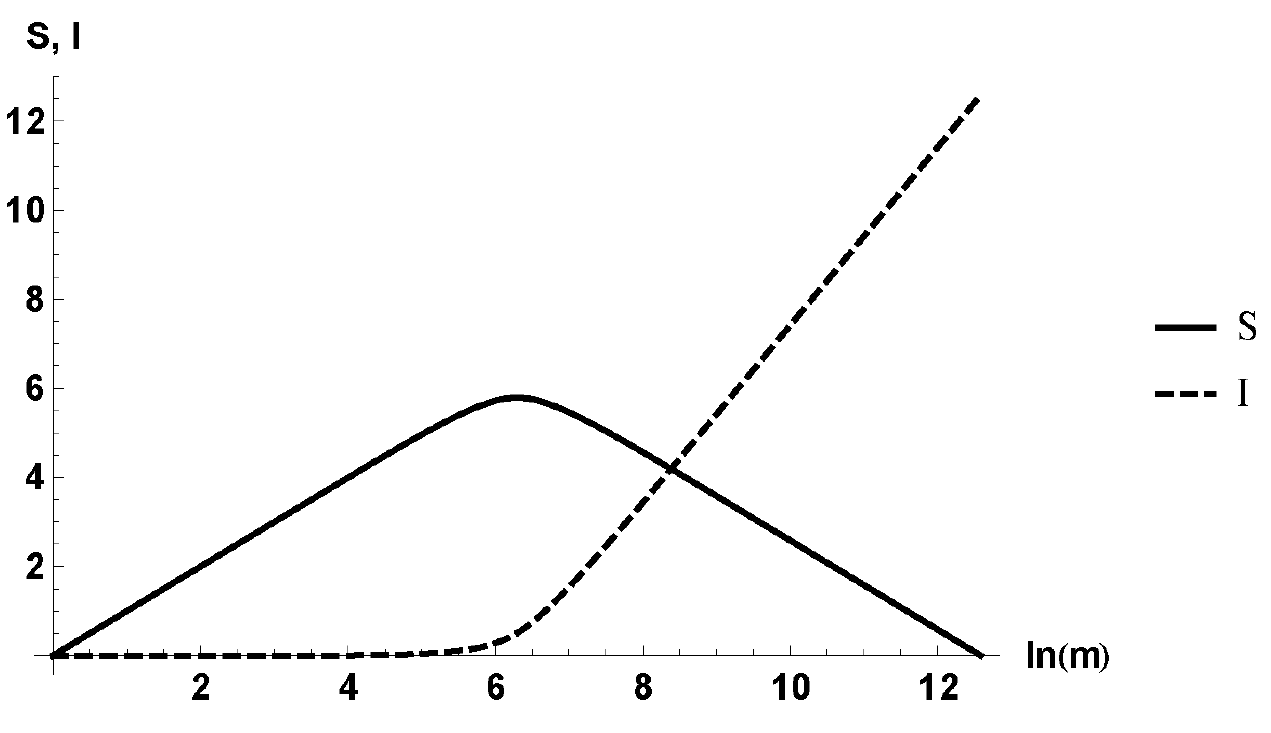}
\caption{ Page entropy of BH radiation (solid) $S_{m,n}$ and information (dashed) $I=\ln m - S_{m,n}$ vs $\ln m$ for $m n = 291,600$ after \cite{Page:1993b}).
The interpretation is that the information in the BH leaks out into the outgoing radiation at roughly half the evaporation (Page) time of the BH.
}\label{Page:S_I_vs_lnm}
\end{center}
\end{figure}
The interpretation is that the information in the BH leaks out into the outgoing radiation at roughly half the evaporation (Page) time of the BH. The initial Hawking radiation (HR) is essentially thermal in nature, while late time HR is non- or nearly- thermal, containing correlations between the BH and emitted radiation. One of the goals of this current work is to develop optically inspired models of BH evaporation capable of reproducing the main features of the Page curve.

\section{An Optically Inspired Model for Unitary BH Evaporation: The Model}\label{sec:model}
\begin{figure}[ht]
\centering
\includegraphics[width=\columnwidth]{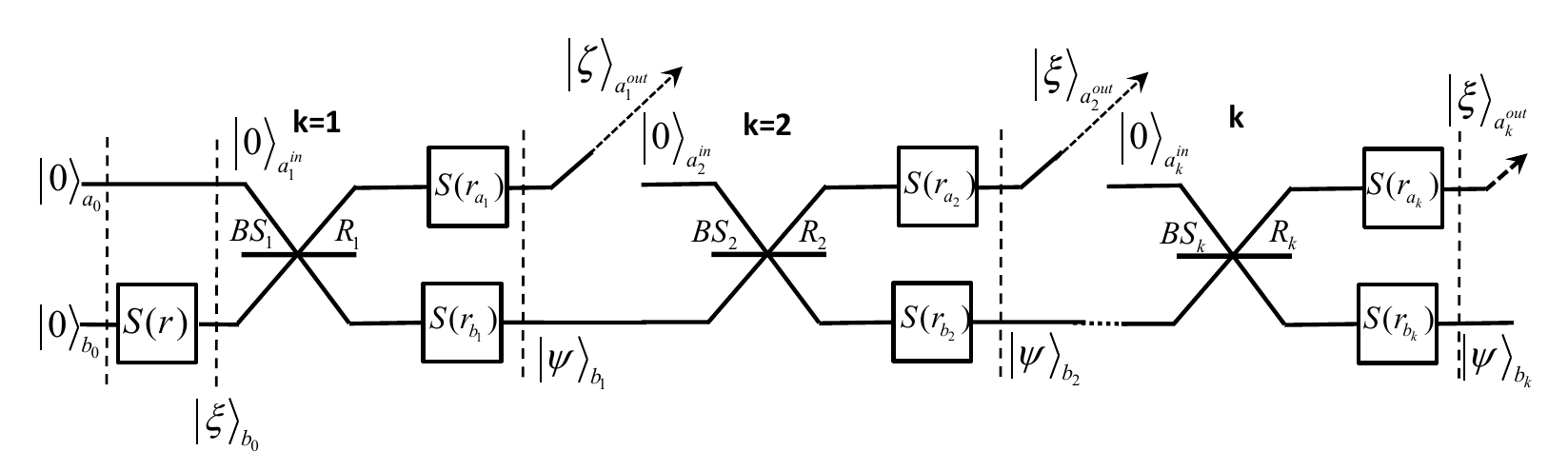}
\caption{BH Evaporation Circuit}
\label{fig:BHEvap:Circuit:plain}
\end{figure}

\noindent We consider a model for unitary black hole vaporation (BHE) as given by the circuit in \Fig{fig:BHEvap:Circuit:plain}
Here the lower rail represents the BH which we put into an initial single-mode squeezed state $\ket{\zeta}_{b_0}\equiv S(r)\ket{0}_{b_0}$ with squeezing parameter $r$ such that $N_{b_0}(0)= \sinh^2 r$. The upper rail represents a succession of initial vacuum modes $\ket{0}_{a_k}$ labeled by the index $k$, which interact with the BH through a beam splitter (BS) interaction with reflectivity $R_k$ that allows a variable swapping of populations between the particular incoming vacuum mode (considered just outside the BH's apparent horizon) and the BH. After the BS interaction, the output of the vacuum mode is unitary transformed to the state $\ket{\xi}_{a_k}$, with mean occupation number $N_{a_k}$.

Since all states are Gaussian states (GS) it is sufficient to work in the Heisenberg picture and simply track the evolution of the quadrature operators
$\Xakoutp= (a^{out}_k)^{\dagger} + a^{out}_k$,  $\Xakoutm= i\,[ (a^{out}_k)^{\dagger} - a^{out}_k]$, and similarly for $\Xbkpm$.
Note that under a BS transformation the quadrature operators at each iteration $k$ transform as
\bea{BS:X:eqns}
\Xakoutpm &=&  \sqrt{\Rk} \, \Xakinpm  + \sqrt{(1-\Rk)}\, \Xbkmpm, \\
\Xbkpm &=&  \sqrt{(1-\Rk)} \, \Xakinpm  -      \sqrt{\Rk}\, \Xbkmpm.
\eea
This immediately leads to the evolution of the variances as
\bea{BS:V:eqns}
\Vakpm &=&  \Rk         + (1-\Rk)\, \Vbkmpm, \label{BS:V:eqns:a}\\
\Vbkpm &=&  (1-\Rk)   +     \Rk\,   \Vbkmpm,\label{BS:V:eqns:b}
\eea
where we have taken all means values to be zero, and have taken $\Vakinpm = \langle (\Xakinpm)^2 \rangle = 1$.

\begin{figure}[htb]
\centering
\includegraphics[width=5.0in,height=3.5in]{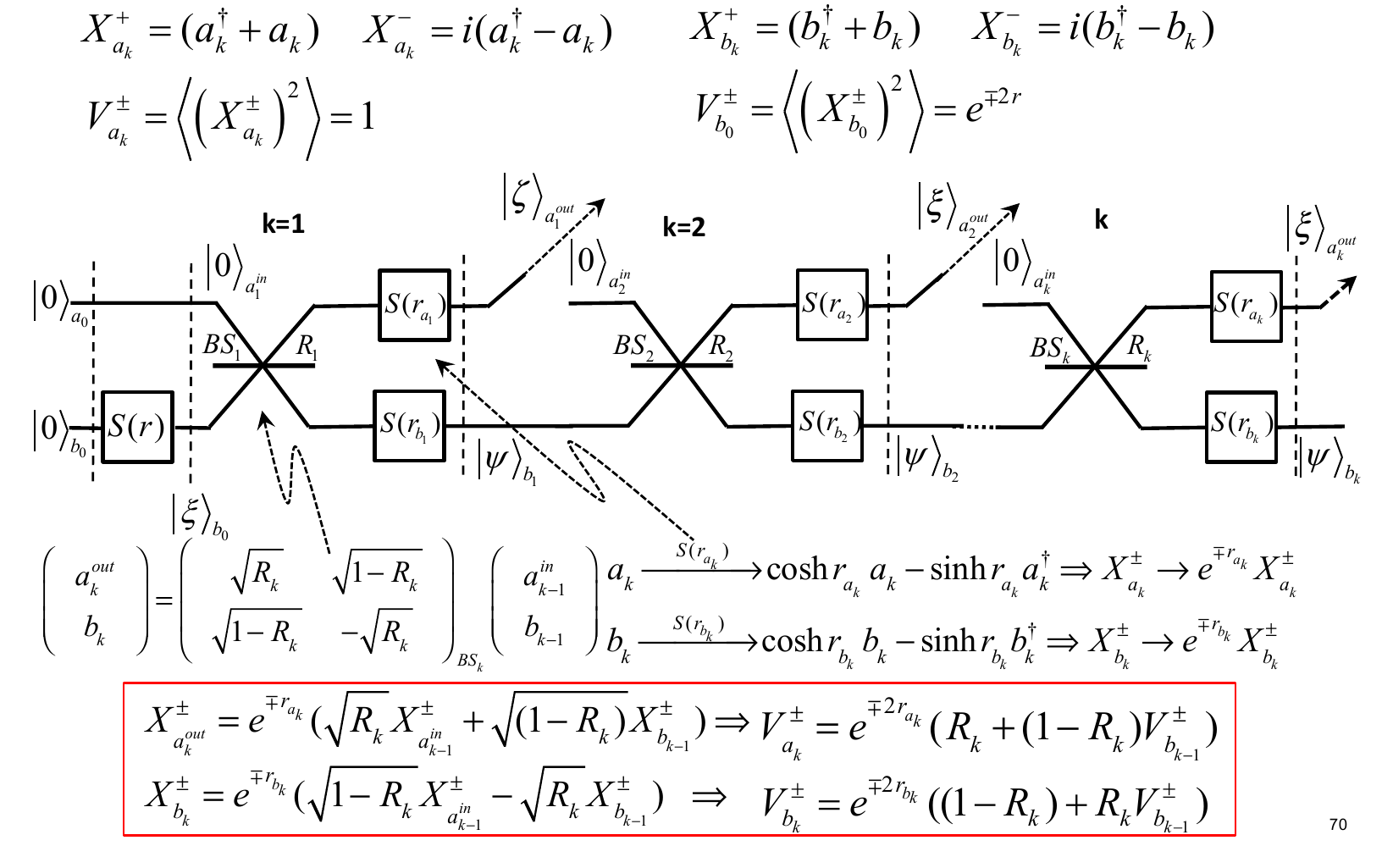}
\caption{Summary of black hole evaporation beam splitter and single mode squeezing for each interaction event labeled by integer $k$.}
\label{fig:BHEvap:Opr:Summary}
\end{figure}

After the BS interaction, the $a^{out}_k$ and $b_k$ each undergo a separate squeezing transformation $S(\rak)$ and $S(\rbk)$ with squeezing parameters $\rak$ and $\rbk$ respectively. The quadratures evolve as 
$\Xakoutpm\rightarrow S(\rak)\,\Xakoutpm\,S^\dagger(\rak) = e^{\mp\rak}\Xakoutpm$, and similarly for $\Xbkpm$.
Hence, the variances evolve simply as
$\Vakpm\rightarrow  e^{\mp 2\,\rak}\Vakpm$,  and $\Vbkpm\rightarrow  e^{\mp 2\,\rbk}\Vbkpm$.
Thus, the full evolution of the quadratures and variances at each interaction/iteration $k$ is given as
\bea{BSSQ:X:eqns}
\Xakoutpm &=& e^{\mp\rak} \left[ \sqrt{\Rk} \, \Xakinpm  + \sqrt{(1-\Rk)}\, \Xbkmpm \right], \label{BSSQ:X:eqns:a}\\
\Xbkpm &=& e^{\mp\rbk} \left[  \sqrt{(1-\Rk)} \, \Xakinpm  -      \sqrt{\Rk}\, \Xbkmpm \right]. \label{BSSQ:X:eqns:b}
\eea
This immediately leads to the evolution of the variances as
\bea{BSSQ:V:eqns}
\Vakpm &=& \erakmp \left[ \Rk         + (1-\Rk)\, \Vbkmpm \right],  \label{BSSQ:V:eqns:a}\\
\Vbkpm &=&  \erbkmp \left[ (1-\Rk)   +     \Rk\,   \Vbkmpm \right]. \label{BSSQ:V:eqns:b}
\eea
The above equations represent the main equations governing our optical analogue BH evaporation model, 
along with the initial conditions $V^\pm_{b_0} \equiv e^{\mp 2\, r_{b_0}} = e^{\mp 2 r}$,
and are illustrated in \Fig{fig:BHEvap:Opr:Summary}.

The purpose of the post-BS squeezing transformations are twofold. First, $S(\rak)$ is invoked on the mode $a^{out}_k$ in order to ensure thermality of the outgoing Hawking Radiation (HR), i.e. we choose $\rak$ to ensure the required condition 
$\Vakp = \Vakm$. 
Secondly, we choose $\rbk$ to ensure conservation of overall occupation number, i.e. $\Nak + \Nbk = \Nbkm$ 
(population exiting BS $k$ and its subsequent squeezer as HR, and the diminished BH, equals the population entering BS $k$ from the BH).
\begin{figure}[htb]
\centering
\includegraphics[width=5.0in,height=3.5in]{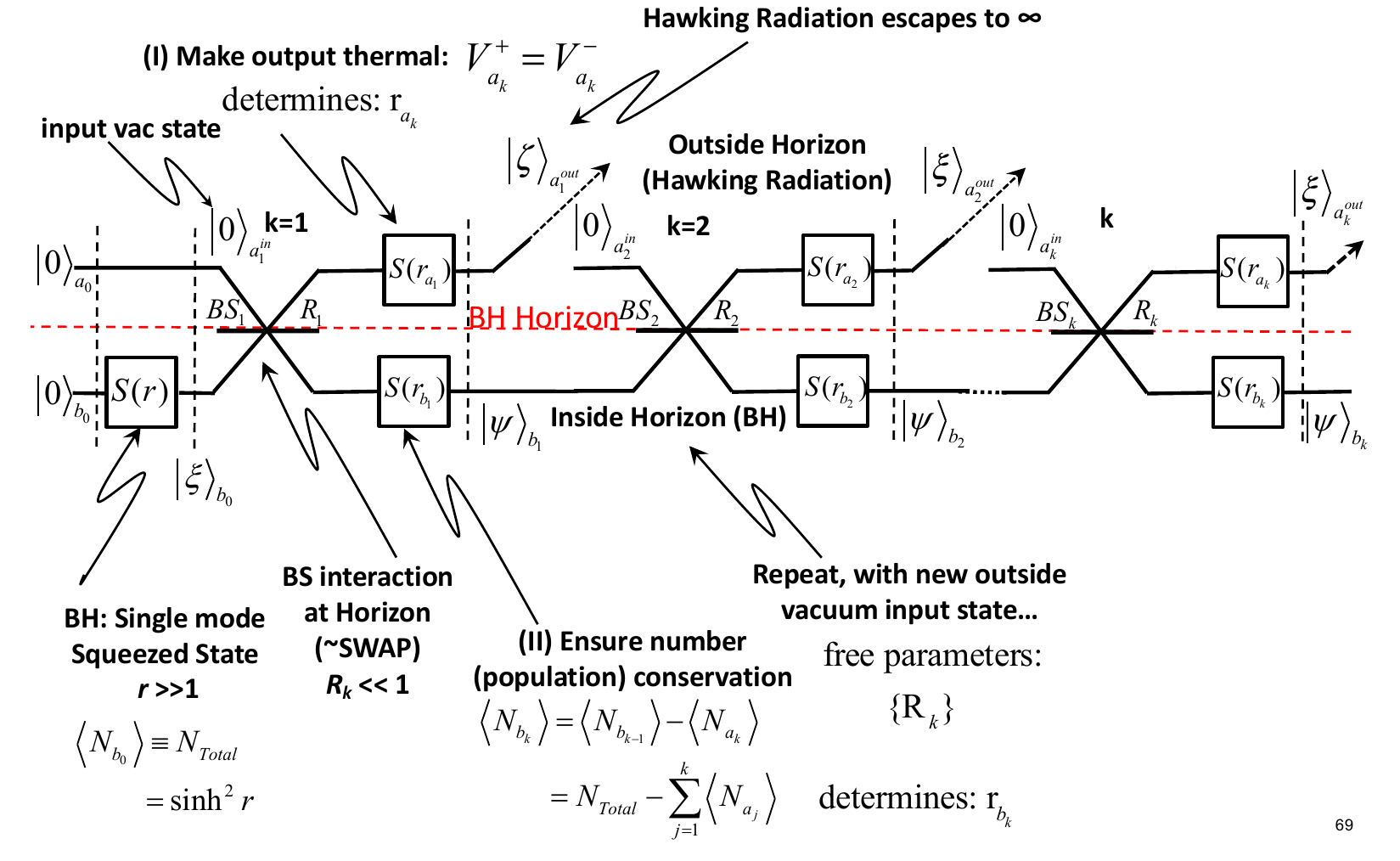}
\caption{Summary of black hole evaporation procedure for each interaction event labeled by integer $k$.}
\label{fig:BHEvap:Procedure}
\end{figure}
The above operational procedure governing our optical analogue BHE model at each interaction labeled by integer $k$ is illustrated in \Fig{fig:BHEvap:Procedure}.
Note that the free parameters in our model is then the set of BS reflectivities $\{R_k\}$, which can chose freely, in order to model certain features of BH evolution.

\section{The role of BS alone}\label{sec:thermal:states}
Before we explore the full BS/Squeezing (BSSQ) model in the previous section, we first look at a simpler case - the role of just the BSs (i.e. no squeezing).
Some key features of BH evaporation we would like to reproduce are (i) the thermality, or near-thermality of the emitted HR ($a^{out}_k$ mode), and (ii) the complete evaporation of the BH (to the vacuum state $\ket{0}_b$) vs evolution to a remnant state).

The BSs with reflectivities $\{R_k\}$ essentially represents a "leaky cavity" whose rate of leakage can be controlled by the parameters $\{R_k\}$,
If $R_k = \epsilon_k$ with $\epsilon_k\ll 1$,  the the BH is almost complete transparent with transmissivity $T=1-R\sim 1$, and there is a very rapid transfer of population from the BH (modes $b_k$) to the outgoing HR (modes $a^{out}_k$). On the other hand, if 
$R_k = 1-\epsilon_k$ the BH is nearly opaque ($T\sim\epsilon_k$), and based on the value of $\epsilon_k$ one can control the occupation number $\Nak$ of the emitted HR at iteration $k$. 

A reasonable choice is simply to explore $R_k\equiv R=1-\epsilon $, i.e. constant for all $k$,
Another reasonable possibility, based on considerations of trying to reproduce the Page Curves, and the fact that $\Nbk$ is taken to be proportional to the mass at iteration $k$ of the evaporating BH, is to explore values of $R_k$ that depend on $\Nbk$. In the following we explore these two cases, (i) $R_k=R=1-\epsilon$ for all $k$ and (ii) $R_k = (\Nbk-R_{\textrm{\scriptsize offset}})/N_{b_0}$ where $R_{\textrm{\scriptsize offset}}\sim\mathcal{O}(1)$ to ensure that $0\le R_k\le 1$.

For the initial BH  we take $\rho_{b_0}$ to be a thermal state with mean occupation number~$\Nb0$
\be{rho:b:0}
\rho_{b_0} = \sum_{n=0}^\infty \frac{(\Nb0)^n}{(1+\Nb0)^{n+1}}\,\ket{n}_b\bra{n},
\ee
and $\rho_{a^{in}_k} = \ket{0}_{a_k}\bra{0}$, both GS (Gaussian States). There is no squeezing so that $\rak=\rbk=0$, and hence
the variance evolve according to \Eq{BS:V:eqns:a} and \Eq{BS:V:eqns:b}. Note that at all times we have 
$\Vakp = \Vakm = 2 \Nak +1$ and $\Vbkp=\Vbkm=2 \Nbk+1$, at each iteration $k$, so that the output HR and the BH remain  thermal state with mean occupation number $\Nak$ and $\Nbk$, respectively. Conservation of overall occupation number $\Nak + \Nbk = \Nbkm$ is assured since the unitary BS transformation is occupation number conserving. During the course of the evolution, no correlations are built up between the BH and the outgoing HR, as each remains in its own thermal states. The thermal BH essentially slowly "leaks out" of the "cavity" to the thermal HR.

\subsection{$R_k\equiv R=1-\epsilon$ for all $k$}
Here we choose $r=2.5$ so that $\Nb0=36.5$. The specific value of $\Nb0$ is not particularly relevant, just as long it is much larger than any $\Nak$ for early times. 
\begin{figure}[ht]
\begin{center}
\begin{tabular}{ccc}
\includegraphics[width=2.5in,height=1.75in]{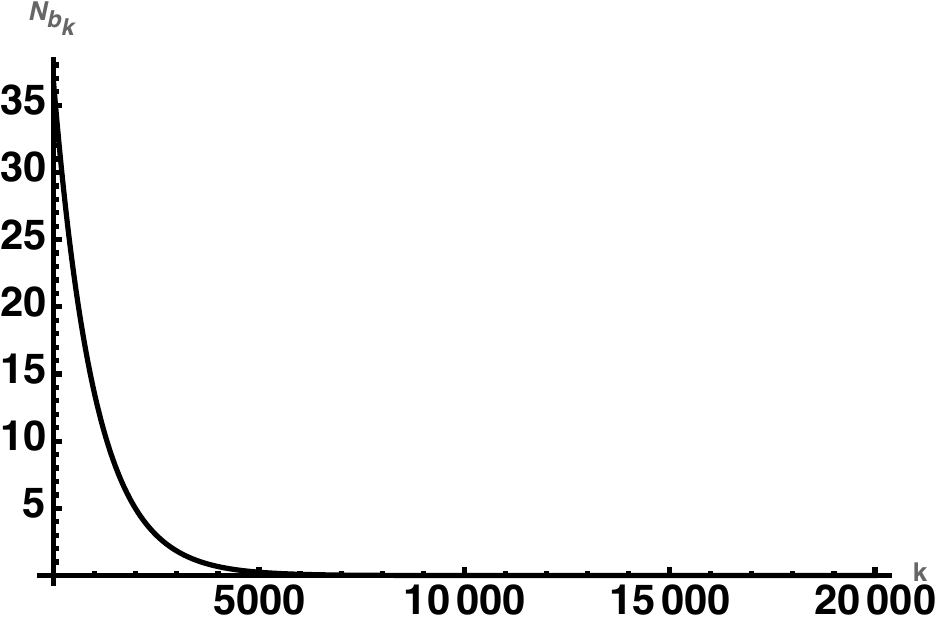} &
{\hspace{0.1in}}
\includegraphics[width=2.5in,height=1.75in]{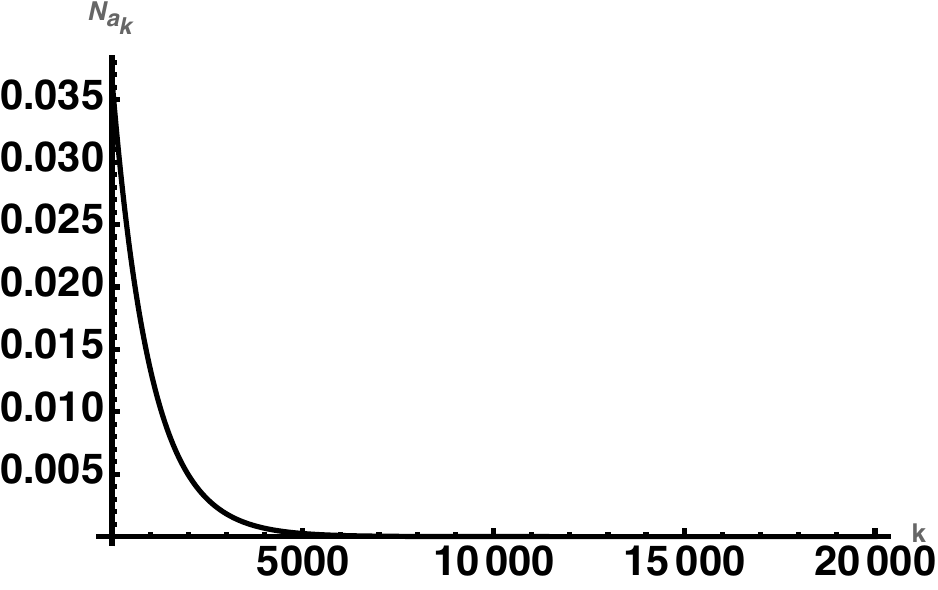}
\end{tabular}
\end{center}
\caption{Occupation numbers ($r=2.5$):  (left) $\Nbk$ (BH) and (right) $\Nak$ (HR) for $R_k = 0.999$ for all $k$.}
\label{fig:Nak:Nbk:R:constant:r:2p5}
\end{figure}
\begin{figure}[ht]
\begin{center}
\begin{tabular}{ccc}
\includegraphics[width=2.5in,height=1.75in]{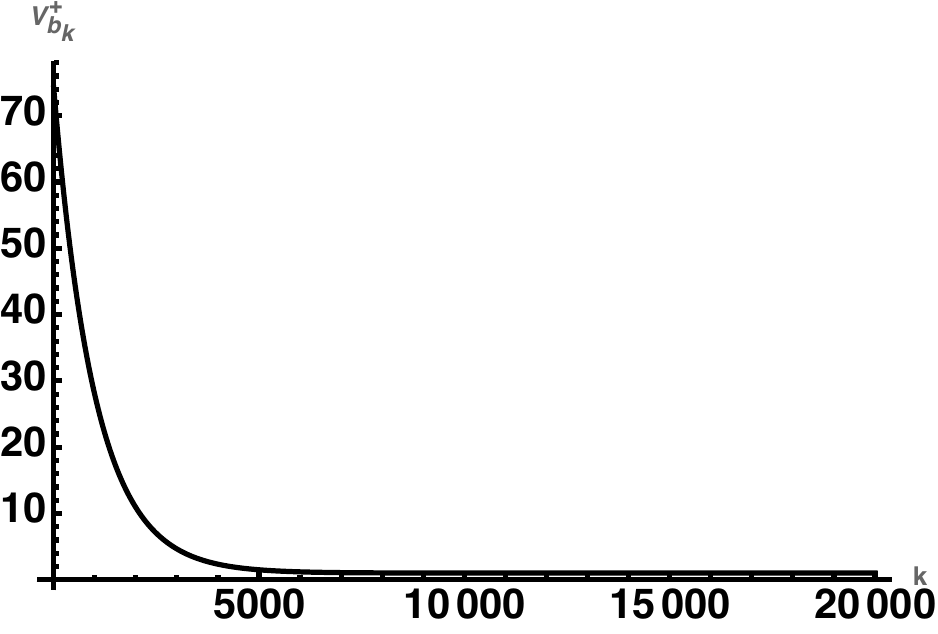} &
{\hspace{0.1in}}
\includegraphics[width=2.5in,height=1.75in]{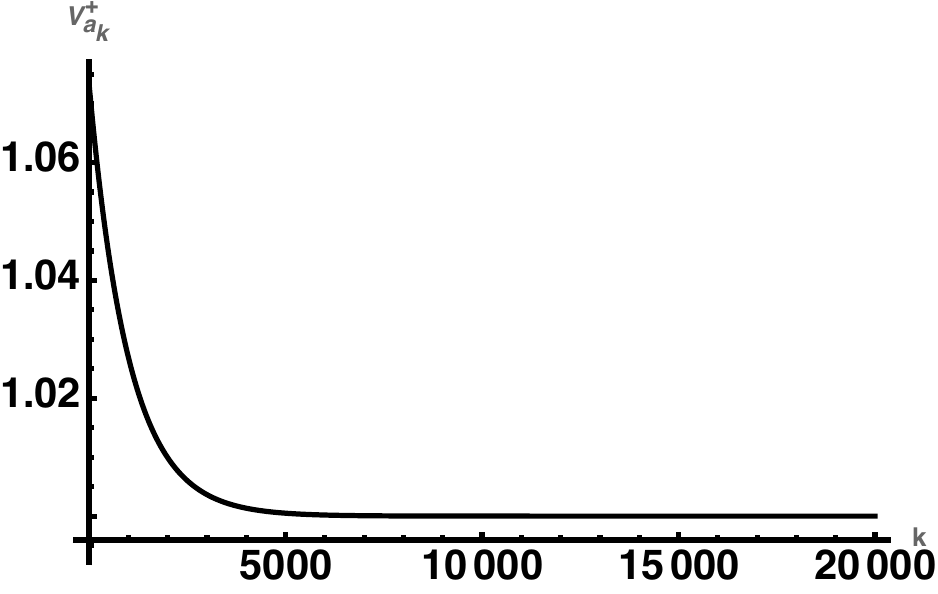}
\end{tabular}
\end{center}
\caption{Variances ($r=2.5$):  (left) $\Vbkp$ (BH) and (right) $\Vakp$ (HR) for $R_k=0.999$ for all $k$.}
\label{fig:Vak:Vbk:R:constant:r:2p5}
\end{figure}
As discussed above, no correlations are developed between the BH and the outgoing HR.
The rate at which the BH radiates is set by the value $\epsilon_k\ll1$ where $R_k=1-\epsilon_k$.  The closer $R_k$ to unity, the slower the thermal BH leaks into the thermal HR.

For $R_k=R=1-\ep$ constant, we can iterate \Eq{BS:V:eqns:a} and \Eq{BS:V:eqns:b} to obtain
\bea{R:constant:iterate}
\hspace{-1.0in}
\Vakpm &=& R\, \sum_{j=0}^{k} (1-R)^j + (1-R)^k\,\Vbzpm  \rightarrow 1 + \ep^k\,\Vbzpm
                      \rightarrow 1 + e^{-k\,|\ln\ep|}\,\Vbzpm\rightarrow 1 + e^{-k\,|\ln\ep| \mp 2\, r}, \qquad \label{R:constant:iterate:a} \\
%
%
\hspace{-1.0in}
\Vbkpm &=& (1-R)\, \sum_{j=0}^{k} R^j+ R^k\, \Vbzpm  \rightarrow 1 + R^k\, \Vbzpm  
= 1 + e^{-k\,|\ln(1-\ep)| \mp 2\,r} \approx 1 + e^{-k\,\ep\, \mp\,2\,r}
\rightarrow 1, \label{R:constant:iterate:b} 
\eea
 employing $0\le R\le 1$, and we have taken 
 $V^\pm_{b_0}  = e^{\mp 2 r}$.
 \Eq{R:constant:iterate:a} and  \Eq{R:constant:iterate:b} 
 imply that $\Vakpm\rightarrow 1$ and $\Vbkpm\rightarrow 1$
  if $k\ge 2\,r/\ep\gg 1$.
Thus, both the BH and the HR eventually evolve the vacuum state with $\Vakp=\Vakm=1=\Vbkp=\Vbkm$, i.e. the BH completely evaporates. The mean populations and variances are shown in 
\Fig{fig:Nak:Nbk:R:constant:r:2p5} and \Fig{fig:Vak:Vbk:R:constant:r:2p5}, respectively for 
$R=0.999$.

\subsection{$R_k = (\Nbk-R_{\textrm{\scriptsize offset}})/N_{b_0}$}
\subsubsection{Case $r=2.5:$}
\begin{figure}[htb]
\centering
\includegraphics[width=2.5in,height=1.75in]{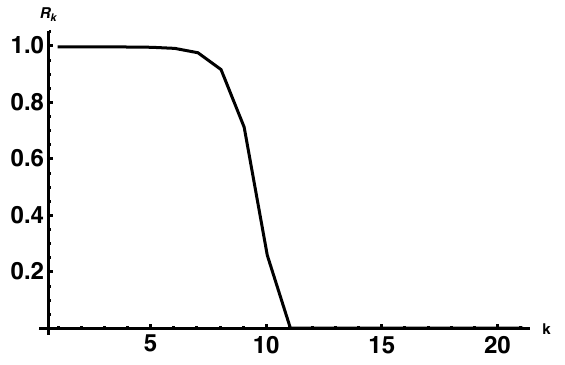}
\caption{$R_k = (\Nbk-R_{\textrm{\scriptsize offset}})/N_{b_0}$ for $r=2.5$.}
\label{fig:Rk:of:Nbk}
\end{figure}

Here we again choose $r=2.5$ so that $\Nb0=36.5$ and make $R_k$ dynamic by setting
\be{Rk:of:Nbk}
R_k = (\Nbk-R_{\textrm{\scriptsize offset}})/N_{b_0}
\ee
and choose $R_{\textrm{\scriptsize offset}}=1$.
The profile of $R_k$ is given in \Fig{fig:Rk:of:Nbk}, which is reminiscent of the behavior of an inverse step function 
$\big(1-\tanh(\beta k)\big)/2$ for some appropriate value of the switching parameter $\beta$.
\begin{figure}[ht]
\begin{tabular}{ccc}
\includegraphics[width=2.5in,height=1.75in]{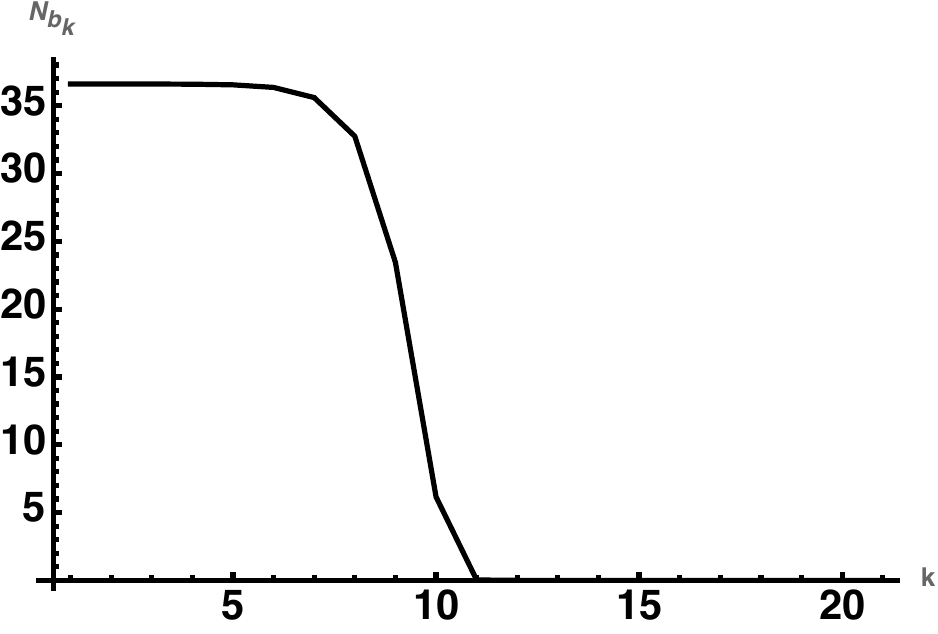} &
{\hspace{0.1in}} &
\includegraphics[width=2.5in,height=1.75in]{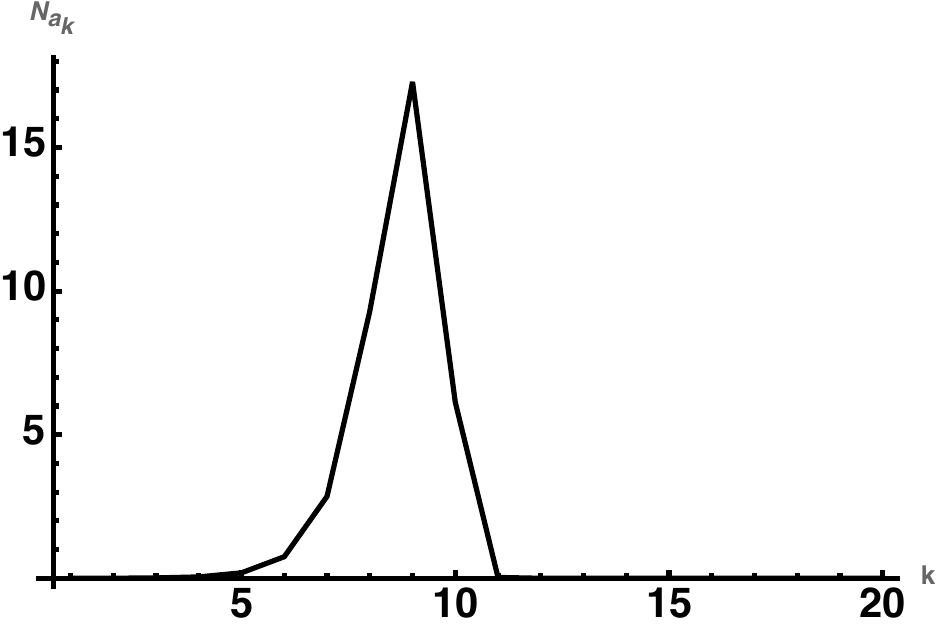}
\end{tabular}
\caption{Occupation numbers: $r=2.5$  (left) $\Nbk$ (BH) and (right) $\Nak$ (HR) for $R_k = (\Nbk-R_{\textrm{\scriptsize offset}})/N_{b_0}$.}
\label{fig:Nak:Nbk:R:dynamic:r:2p5}
\end{figure}
What is more interesting is the "bursting-like" behavior which occurs towards the end of the BH's life as shown in \Fig{fig:Nak:Nbk:R:dynamic:r:2p5}. Here we see $\Nak$ approximately zero for most of the BH life, but with a sudden rapid rise and decrease in the occupation number towards the end as $\Nbk$ suddenly and rapidly plummets towards zero. 
\begin{figure}[ht]
\begin{tabular}{ccc}
\includegraphics[width=2.5in,height=1.75in]{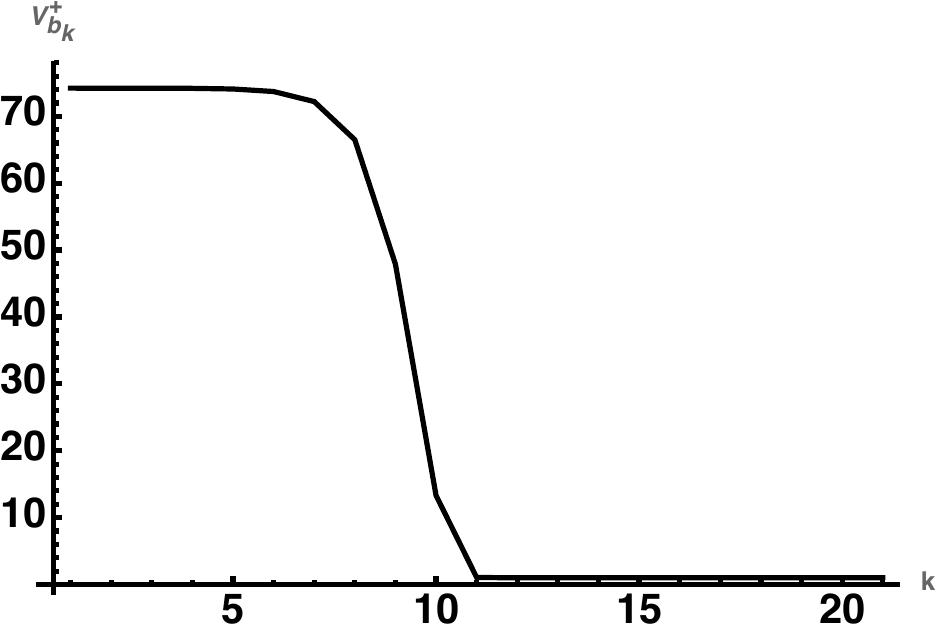} &
{\hspace{0.1in}} &
\includegraphics[width=2.5in,height=1.75in]{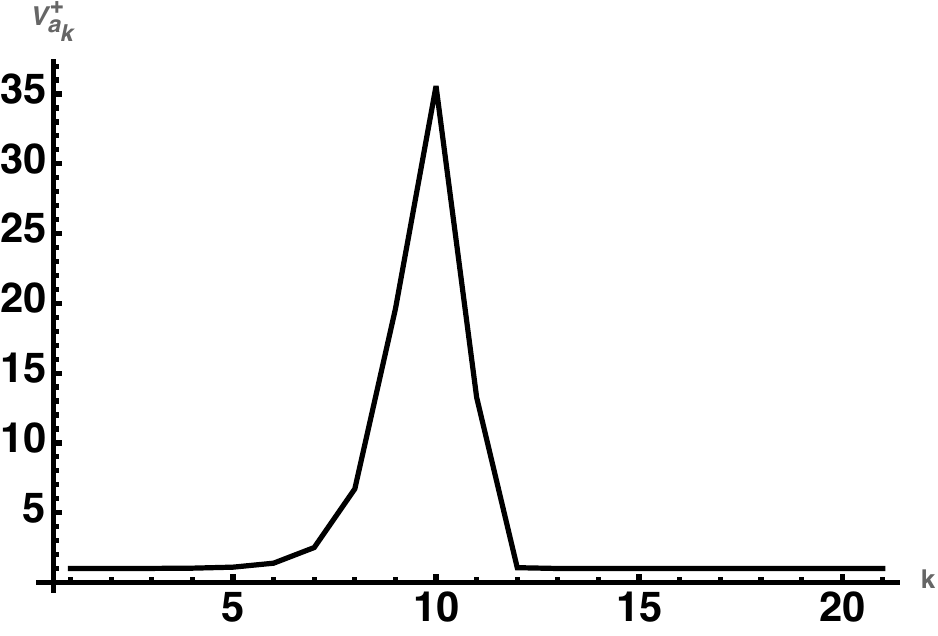}
\end{tabular}
\caption{Variances ($r=2.5$):  (left) $\Vbkp$ (BH) and (right) $\Vakp$ (HR) for $R_k = (\Nbk-R_{\textrm{\scriptsize offset}})/N_{b_0}$.}
\label{fig:Vak:Vbk:R:dynamic:r:2p5}
\end{figure}
Since in this thermal state case, the variances are just twice the occupation number plus unity, the same behavior is reflected in $\Vbkp$ and $\Vakp$ in \Fig{fig:Vak:Vbk:R:dynamic:r:2p5}.

\subsubsection{Case $r=7.5:$}
\begin{figure}[htb]
\centering
\includegraphics[width=2.5in,height=1.75in]{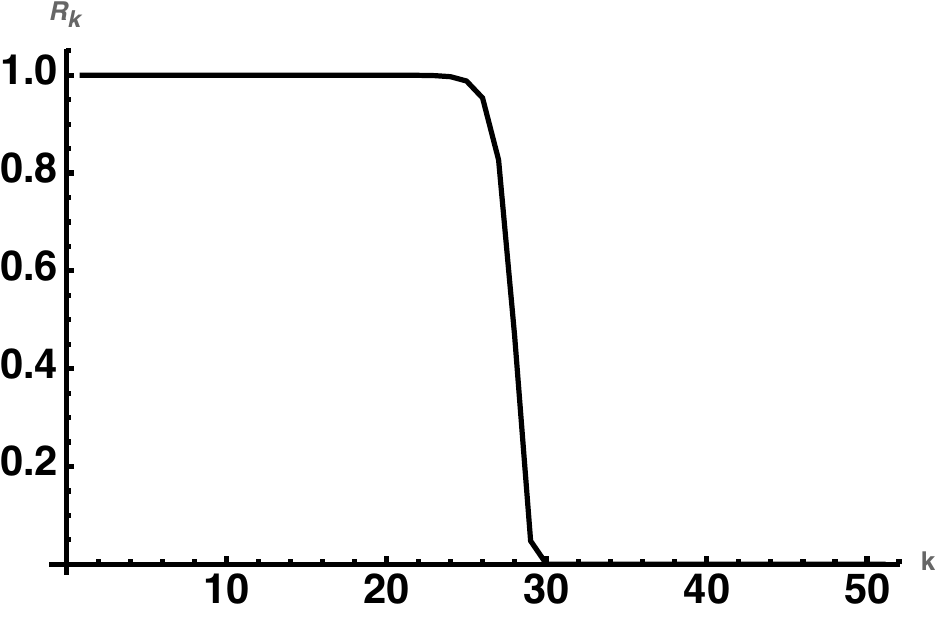}
\caption{$R_k = (\Nbk-R_{\textrm{\scriptsize offset}})/N_{b_0}$ for $r=7.5$.}
\label{fig:Rk:of:Nbk:r:7p5}
\end{figure}

We can smooth out the above figures by starting with a larger $\Nb0=\sinh^2 r$. 
Here we investigate $r=7.5$ so that $\Nb0 =817,254$.
and again choose $R_{\textrm{\scriptsize offset}}=1$
The profile of $R_k$ is given in \Fig{fig:Rk:of:Nbk:r:7p5}, which has the same qualitative behavior as for $r=2.5$.
\begin{figure}[ht]
\begin{center}
\begin{tabular}{ccc}
\includegraphics[width=2.5in,height=1.75in]{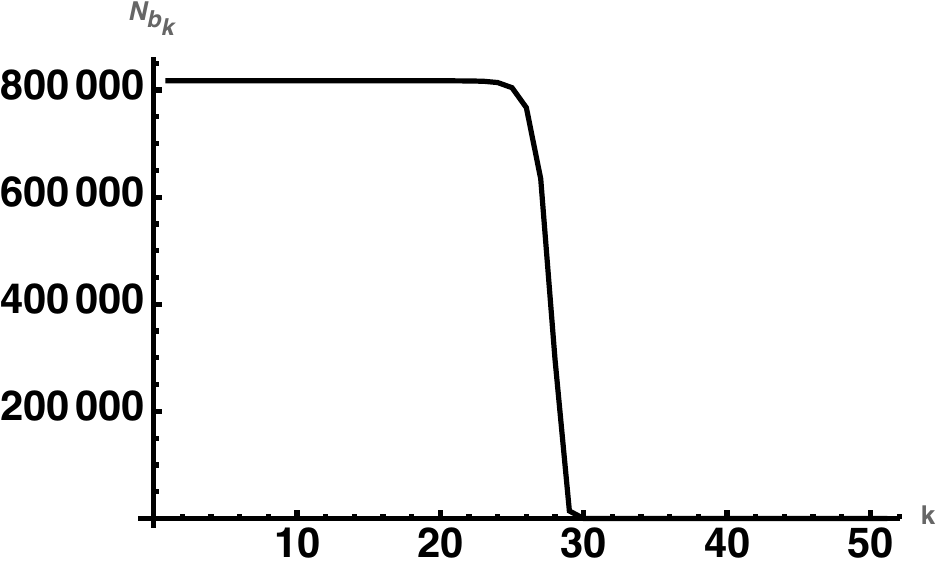} &
{\hspace{0.1in}} &
\includegraphics[width=2.5in,height=1.75in]{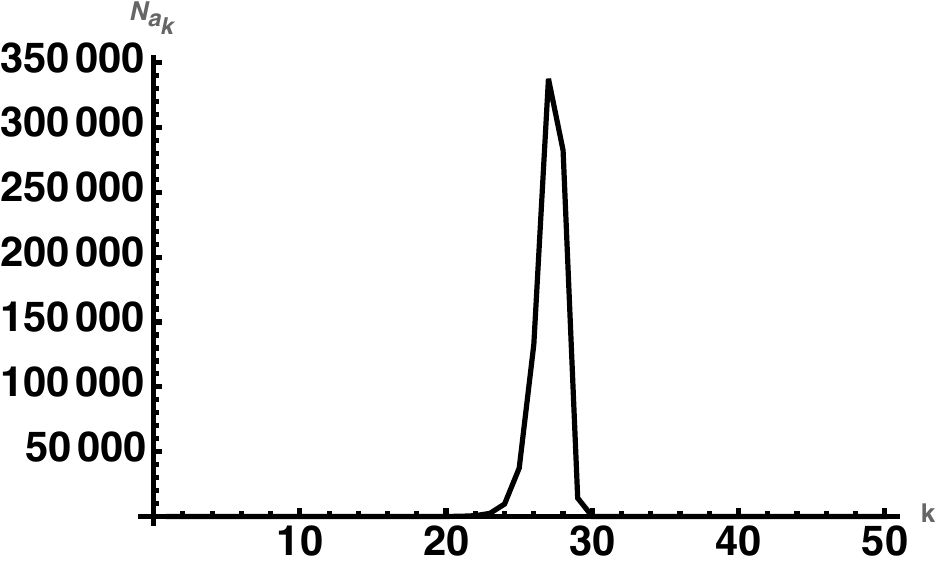}
\end{tabular}
\end{center}
\caption{Occupation numbers: $r=7.5$  (left) $\Nbk$ (BH) and (right) $\Nak$ (HR) for $R_k = (\Nbk-R_{\textrm{\scriptsize offset}})/N_{b_0}$.}
\label{fig:Nak:Nbk:R:dynamic:r:7p5}
\end{figure}
The bursting behavior in \Fig{fig:Nak:Nbk:R:dynamic:r:7p5} reveals that $\Nak$ peaks to roughly $\Nb0/2$.
This runaway process in indicative of a negative heat capacity.
\begin{figure}[ht]
\begin{tabular}{ccc}
\includegraphics[width=2.5in,height=1.75in]{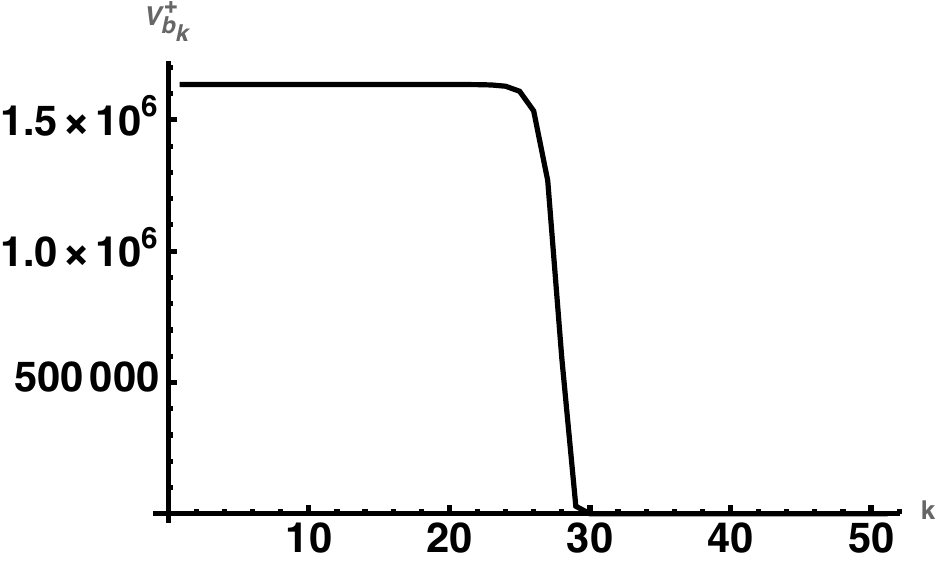} &
{\hspace{0.1in}} &
\includegraphics[width=2.5in,height=1.75in]{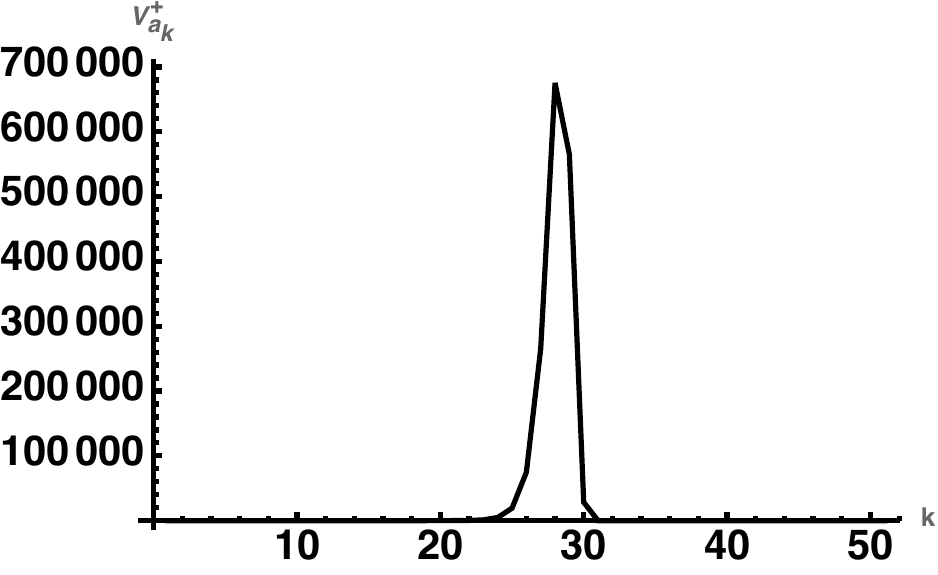}
\end{tabular}
\caption{Variances ($r=7.5$):  (left) $\Vbkp$ (BH) and (right) $\Vakp$ (HR) for $R_k = (\Nbk-R_{\textrm{\scriptsize offset}})/N_{b_0}$.}
\label{fig:Vak:Vbk:R:dynamic:r:7p5}
\end{figure}
Again, for thermal states, the same behavior is reflected in the variances $\Vbkp$ and $\Vakp$ in \Fig{fig:Vak:Vbk:R:dynamic:r:7p5}.

\section{The full model: BSSQ: BS plus Squeezing with $R_k = (\Nbk-R_{\textrm{\scriptsize offset}})/N_{b_0}$ }\label{sec:BS:and:SQ}
\subsection{Occupation numbers and variances}
We now explore the full model governed by \Eq{BSSQ:V:eqns:a} and \Eq{BSSQ:V:eqns:b} and illustrated in \Fig{fig:BHEvap:Circuit:plain} that included both the BH/HR BS interaction and the single mode squeezing operations to ensure (i) thermality of the outgoing HR and (ii) conservation of occupation number:
\bea{squeezing:conditions}
(i)\;\Vakp &=& \Vakm, \label{squeezing:conditions:i}\\
(ii)\; \Nbkm &=& \Nbk + \Nak. \label{squeezing:conditions:ii}
\eea
We note the operator identities 
\bea{opr:identities}
\Nak &\equiv& \frac{1}{2}\,\left( \frac{1}{2}\,\left[  \Vakp + \Vakm \right] - 1  \right)  = \frac{1}{2}\,\left( \Vakp  - 1  \right) , \label{opr:identities:a} \\
\Nbk &\equiv& \frac{1}{2}\,\left( \frac{1}{2}\,\left[  \Vbkp + \Vbkm \right] - 1  \right), \label{opr:identities:b}
\eea
where the second equality in \Eq{opr:identities:a} follows from use of condition (i) \Eq{squeezing:conditions:i}.

\begin{figure}[htb]
\centering
\includegraphics[width=2.5in,height=1.75in]{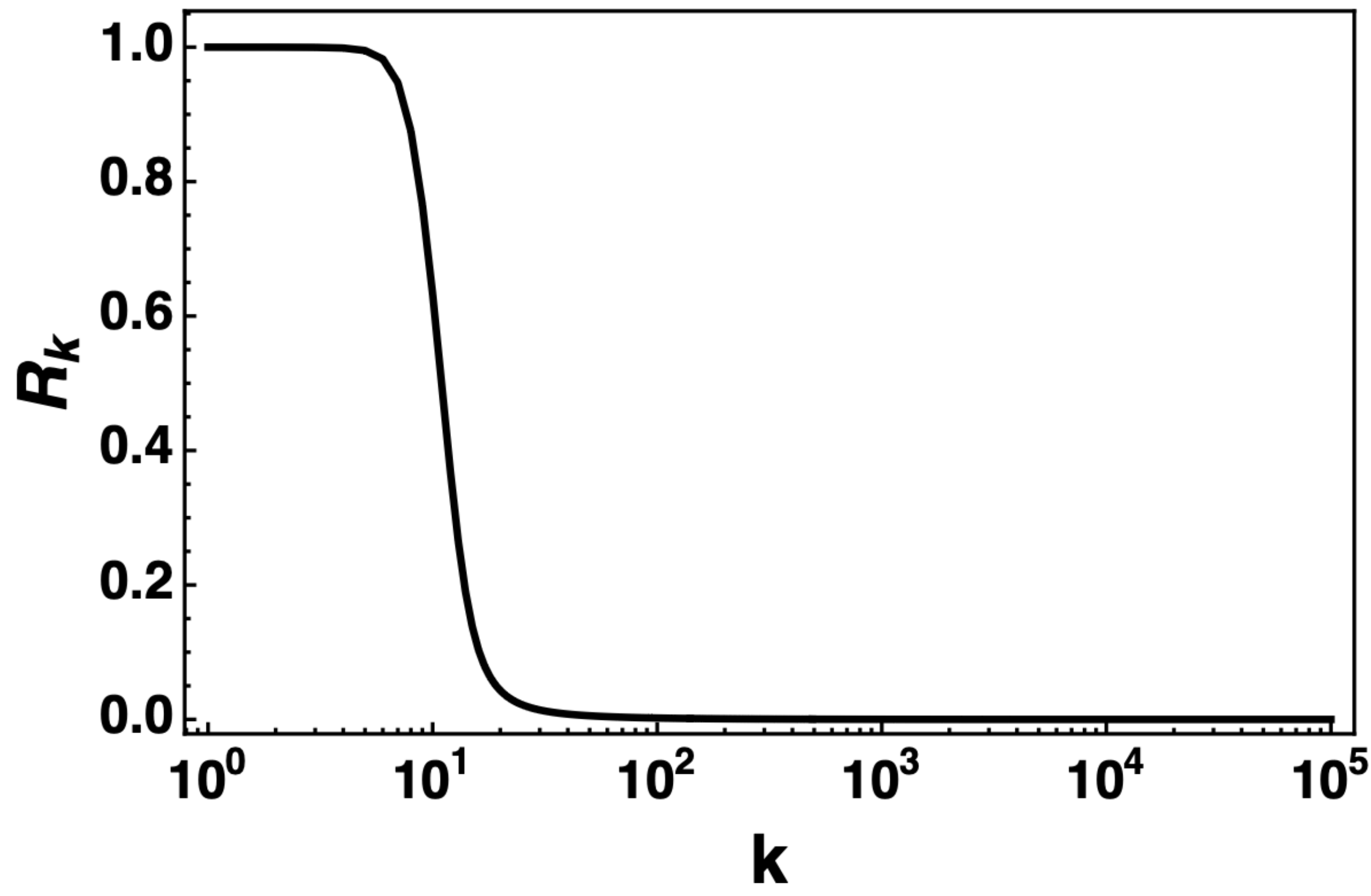}
\caption{$R_k = (\Nbk-R_{\textrm{\scriptsize offset}})/N_{b_0}$ for $r=2.5$ 
for the full model of BS and single mode squeezers, \Eq{BSSQ:V:eqns:a} and \Eq{BSSQ:V:eqns:b}.
}
\label{fig:Rk:of:Nbk:r:2p5:BSSQ}
\end{figure}

\begin{figure}[ht]
\begin{tabular}{ccc}
\includegraphics[width=2.5in,height=1.75in]{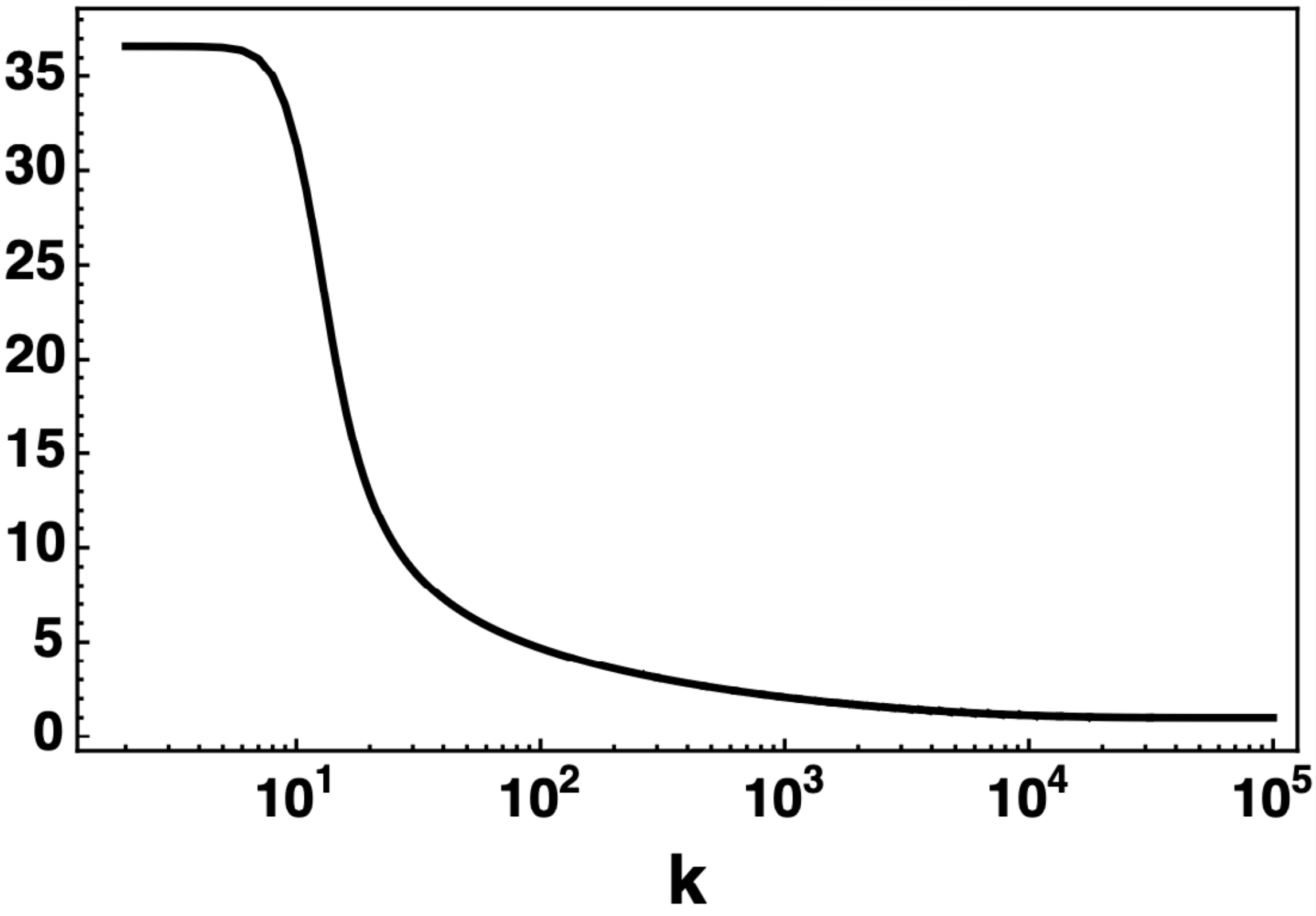} &
{\hspace{0.1in}} &
\includegraphics[width=2.5in,height=1.75in]{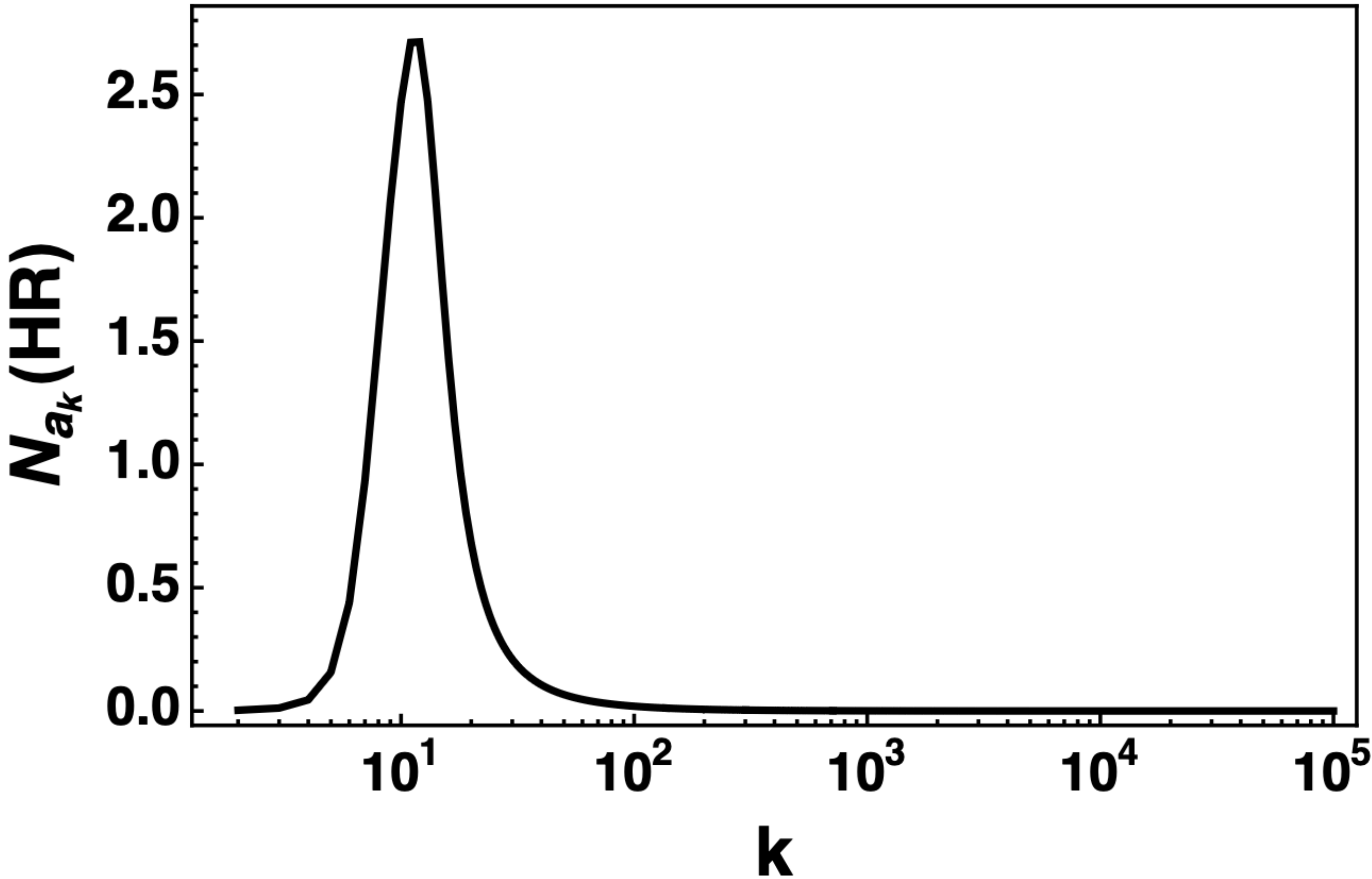}
\end{tabular}
\caption{Occupation numbers: ($r=2.5$):  (left) $\Nbk$ (BH) and (right) $\Nak$ (HR) with $R_k = (\Nbk-R_{\textrm{\scriptsize offset}})/N_{b_0}$ 
for the full model of BS and single mode squeezers, \Eq{BSSQ:V:eqns:a} and \Eq{BSSQ:V:eqns:b}.
}
\label{fig:Nak:Nbk:R:dynamic:r:2p5:BSSQ}
\end{figure}

\Eq{squeezing:conditions:i} is an equation for $\rak$, while \Eq{squeezing:conditions:ii} is then an equation for $\rbk$, both with $R_k$ as a free parameter.
The solutions are given in the Appendix A. For now we examine the case $R_k = (\Nbk-R_{\textrm{\scriptsize offset}})/N_{b_0}$ explored in the previous section (there, without squeezing). We again chose $r=2.5$ and take $R_{\textrm{\scriptsize offset}}=1$.
The profile of $R_k$ is given in \Fig{fig:Rk:of:Nbk:r:2p5:BSSQ}, which has the same qualitative behavior the previous BS-only plots for $r=2.5$.
The BH and HR occupations numbers $\Nbk$ and $\Nak$, respectively, are shown in \Fig{fig:Nak:Nbk:R:dynamic:r:2p5:BSSQ}.
While $\Nbk$ again exhibits a rapid decrease as $R_k$ rapidly transitions from near unity to near zero values, it now has a much slower decaying tail for large values of $k$. We also see a difference in $\Nak$. As in the BS-only case, there is again a "bursting" phenomena, but now it tempered with $\Nak$ never increasing much over the value of $2.5$ and staying very near zero  (i.e. near thermal) for most values of $k$, especially large values of $k$.

\begin{figure}[ht]
\begin{tabular}{ccc}
\includegraphics[width=2.5in,height=1.75in]{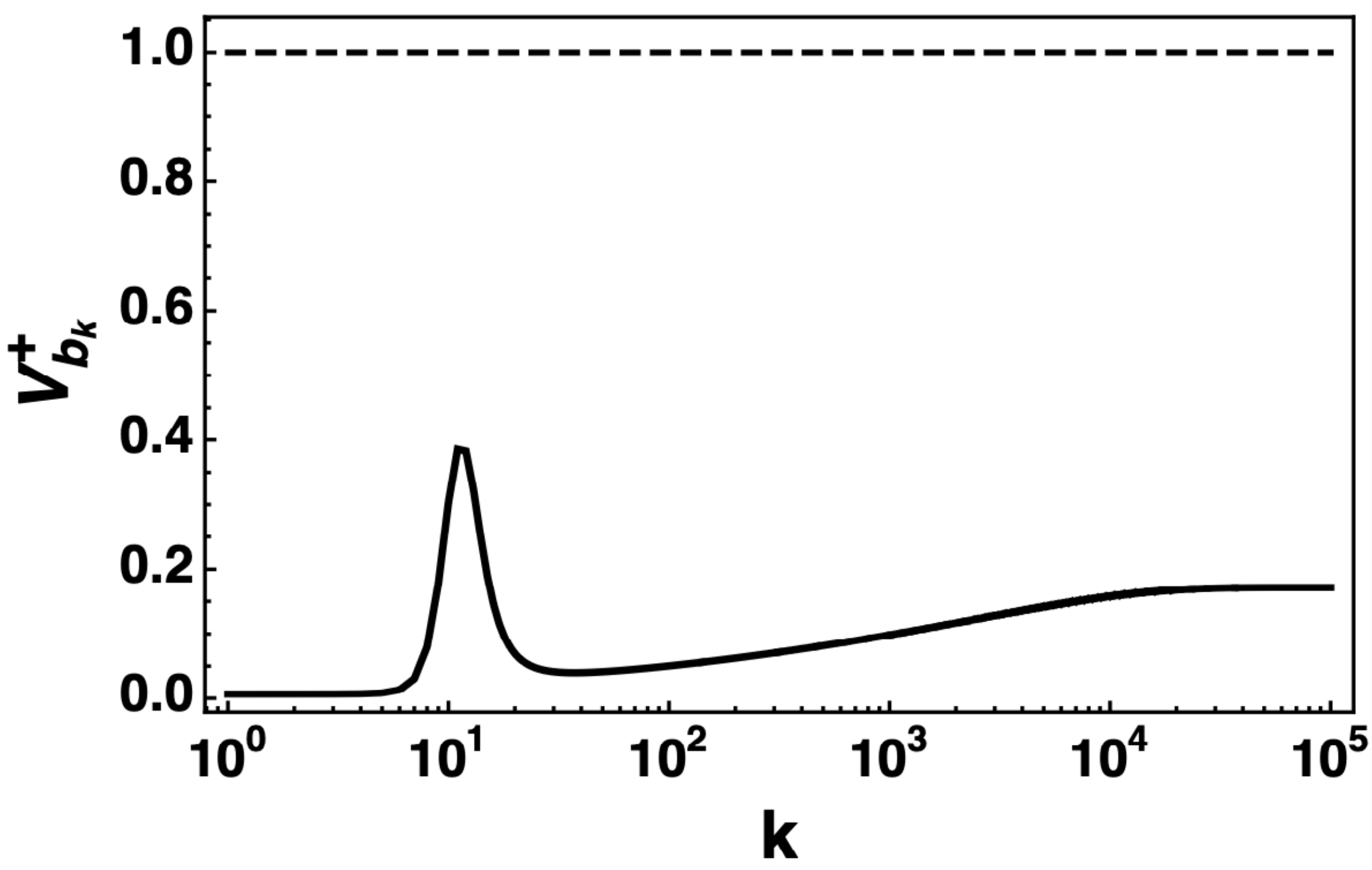} &
{\hspace{0.1in}} &
\includegraphics[width=2.5in,height=1.75in]{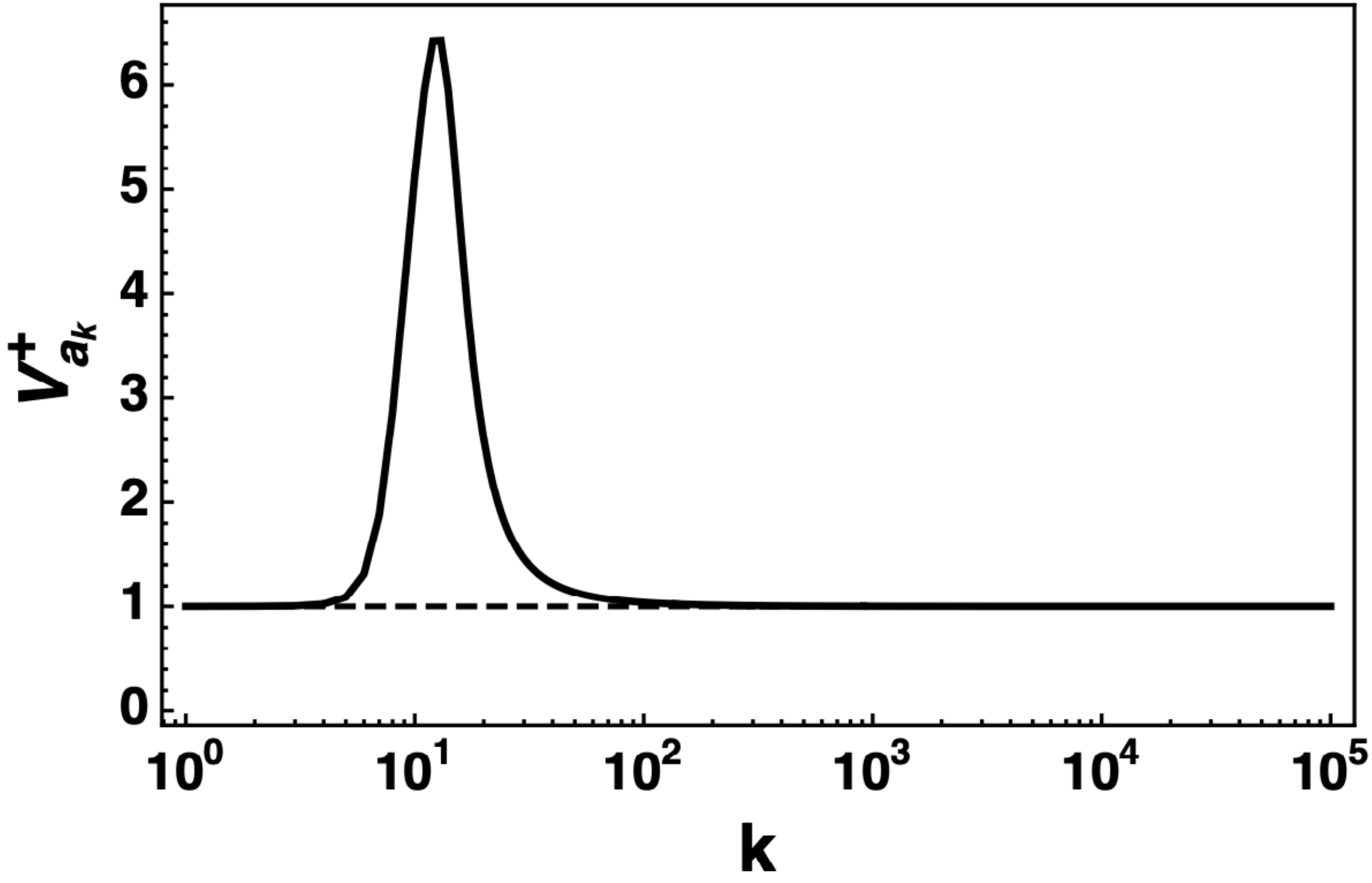}
\end{tabular}
\caption{Variances ($r=2.5$):  (left) $\Vbkp$ (BH) and (right) $\Vakp$ (HR) with $R_k = (\Nbk-R_{\textrm{\scriptsize offset}})/N_{b_0}$ 
for the full model of BS and single mode squeezers, \Eq{BSSQ:V:eqns:a} and \Eq{BSSQ:V:eqns:b}.
}
\label{fig:Vakp:Vbkp:R:dynamic:r:2p5:BSSQ}
\end{figure}

The variances $\Vbkp$ and $\Vakp$ are shown in \Fig{fig:Vakp:Vbkp:R:dynamic:r:2p5:BSSQ}.
Recall that $V^{+}_{b_{0}} = e^{-2\,r} \le 1, (r>0)$ and $V^{+}_{a^{in}_{K}} = 1$.
In \Fig{fig:Vakp:Vbkp:R:dynamic:r:2p5:BSSQ}(left) we see that $\Vbkp\le 1$ for all $k$ and rises above approximately $0.4$.
On the other hand, $\Vakp\ge 1$ for all values of $k$ and peaks to a value of nearly $6.4$ which is nearly twice plus unity of its population $\Nak$ at that iteration, namely $\Vakp\sim 2\,\Nak+1$ at its peak value, that is approximately a thermal state of occupation number $\Nak$.

\begin{figure}[htb]
\centering
\includegraphics[width=2.5in,height=1.75in]{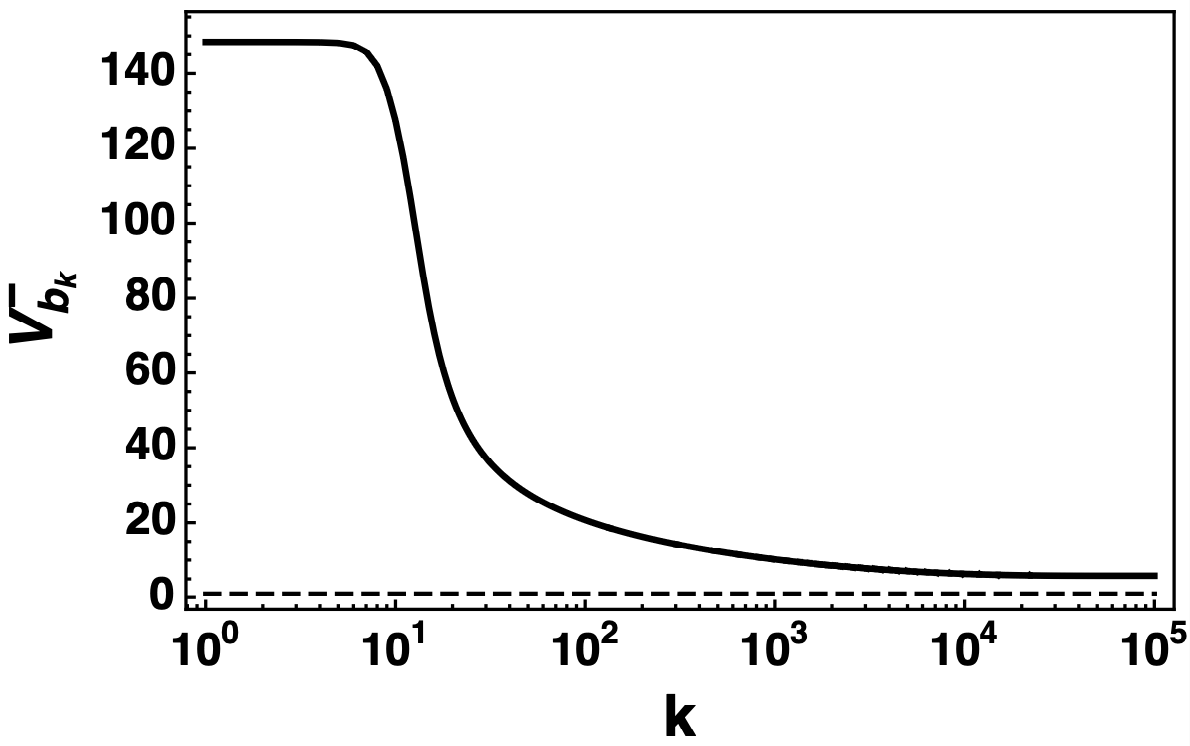}
\caption{($r=2.5$): for  $\Vbkm$ with $R_k = (\Nbk-R_{\textrm{\scriptsize offset}})/N_{b_0}$ 
for the full model of BS and single mode squeezers, \Eq{BSSQ:V:eqns:a} and \Eq{BSSQ:V:eqns:b}.
}
\label{fig:Vbkm:R:dynamic:r:2p5:BSSQ}
\end{figure}

Lastly, in \Fig{fig:Vbkm:R:dynamic:r:2p5:BSSQ} we plot $\Vbkm$. Recall that $\Vbzkm = e^{2\,r} \ge 1$ and we see that $\Vbkm$ follows a profile qualitatively similar to that of $\Rk$ and $\Nbk$.
The message of \Fig{fig:Vakp:Vbkp:R:dynamic:r:2p5:BSSQ} and  \Fig{fig:Vbkm:R:dynamic:r:2p5:BSSQ} is that as $k$ increases the BH evolves towards the vacuum state $\Vbkp=\Vbkm=1$ with $\Vbkp$ approaching unity from above  and $\Vbkm$ from below. The process of BH evaporation in this model, "un-squeezes" (thermalizes) the BH, while the HR is emitted in nearly thermal states of small occupation numbers during the whole process.

In the next section we examine the correlations that are built up between the BH and the emitted HR as it evaporates.

\subsection{BH-HR correlations: Purity and Log Negativity}
To explore the entanglement properties of the full model discussed in the previous section, we utilize the extensive machinery developed for continuous variable Gaussian States (GSs) \cite{Adesso:Illuminati:2007,Weedbrook_GS_RMP:2012,Tserkis_Ralph:2017,Serafini:book:2017}. The most important quantities that characterize the quantum correlations in a multimode GS are its \textit{symplectic eigenvalues} $\nu_j$ for each mode $j$. Symplectic transformations arise as the transformation properties of quantum canonical variables (e.g. $\{x_i, p_i\}$ or equivalently of quadrature variables
$\{X^{+}, X^{-}\}$). The symplectic eigenvalues play a central role since any $n$-mode GS can be written as~\cite{Serafini:book:2017}
\be{rho:G:general}
\rho_G = D^\dagger_{\vec{r}}\, S^\dagger 
\left( 
\otimes_{j=1}^{n}
\left(
\sum_{m=0}^\infty \frac{2}{\nu_j+1} \left( \frac{\nu_j-1}{\nu_j+1}\right)^m\, \ket{m}_j\bra{m}
\right)
\right)
S\,D_{\vec{r}},
\ee
where $D_{\vec{r}}$ are bosonic displacement operators (here $\vec{r} = \{x_1,p_1,x_2,p_2,\ldots,x_n,p_n\}$), 
and $S$ are squeezing transformations. The symplectic eigenvalues can be written as 
$\nu_j = 2\bar{n}_j+1\ge 1$ which converts the term in the inner large parenthesis in  \Eq{rho:G:general} into a thermal state with a mean occupation number 
$\bar{n}_j = 1/\big(\exp(\hbar\omega_j/(k_B T)) -1\big)$ in mode j. 
The implication of \Eq{rho:G:general} is that all GSs are at most quadratic transformations (e.g. BSs and squeezers) of some base multi-mode thermal state. Consequently, it can be shown that the covariance matrix $\bsig$ which characterizes the GS can be written in block diagonal form as
\be{sigma:GS}
\bsig = S\, \Big( \oplus_{j=1}^n \nu_j \Id_2 \Big)\, S^T,
\ee
where $\Id_2$ is the $2\times 2$ identity matrix with doubly degenerate eigenvalues $\nu_j$ for the $j$th 
quadrature mode $\{X_j^{+}, X_j^{-}\}$. Thus, many quantum correlations quantities of interest can be expressed directly in terms of the symplectic eigenvalues $\nu_j$. For example, the
 \textit{purity}, $\mu_G$ can be expressed as 
 \be{purity}
 \mu_G = Tr[\rho_G^2] = \frac{1}{\textrm{Det}\,\bsig} = \Pi_{j=1}^n \frac{1}{\nu_j},
 \ee
which is a measure of the \textit{mixedness} of the quantum states.
The (von Neumann) \textit{entropy}, $S_V(\rho_G)$ of the Gaussian state $\rho_G$ can be expressed as
\bea{entropy}
S_V(\rho_G) &=& -Tr[\rho_G\,\log_2\,\rho_G] = \sum_{j=1}^n s_V(\nu_j), \\
s_V(x) &\equiv& \frac{x+1}{2}\,\log_2\,\left( \frac{x+1}{2}\right) 
                    - \frac{x-1}{2}\,\log_2\,\left( \frac{x-1}{2}\right), \no
{} &=& (\bar{n}_j + 1) \log_2\,( \bar{n}_j + 1) -   \bar{n}_j \, \log_2\,\bar{n}_j. \nonumber                 
\eea
Finally, the \textit{log negativity} $E_{\mathcal{N}} (\rho_G)$ is an bipartite entanglement monotone (vs a \textit{measure}) that is fairly easy to compute and witnesses bipartite entanglement (but in some cases does not detect \textit{bound} entanglement). Essentially, it based on the non-positivity of the partial transpose of $\rho_G$ on one of the two subsystems. The procedure for computing the log negativity is given as follows \cite{Adesso:Illuminati:2007,Serafini:book:2017}.
For a bipartite GS with covariance matrix that can be written as
\be{sig:A:B}
\bsig = 
\left(
\begin{array}{cc}
\bsig_A & \bsig_{A B} \\
\bsig^T_{A B} & \bsig_B\end{array}
\right)
\ee
(i) determine the symplectic invariants Det$\bsig$ and 
$\tilde{\Delta} \equiv  \textrm{Det}\,\bsig_A + \textrm{Det}\,\bsig_B - 2\textrm{Det}\,\bsig_{AB}$
(Note: $\Delta \equiv  \textrm{Det}\,\bsig_A + \textrm{Det}\,\bsig_B + 2\textrm{Det}\,\bsig_{AB}$ is an invariant of $\bsig$, while $\tilde{\Delta}$ is an invariant of its partial transpose), 
(ii) determine the smallest symplectic eigenvalue $\tilde{\nu}_{-}$ of the partial transpose given by
\be{tilde:nu:minus}
\tilde{\nu}^2_{\mp} =\frac{\tilde{\Delta} \mp \sqrt{ \tilde{\Delta} - 4 \textrm{Det}\bsig}}{2},
\ee
and 
(iii) compute the log negative as
\be{log:Neg}
E_{\mathcal{N}} (\rho_G) = \textrm{max} \{0,-\log_2\,\tilde{\nu}_{-} \}.
\ee

\subsubsection{Quantum Correlation Plots}\label{sec:ent:plots}
\begin{figure}[ht]
\begin{center}
\begin{tabular}{ccc}
\includegraphics[width=2.5in,height=1.75in]{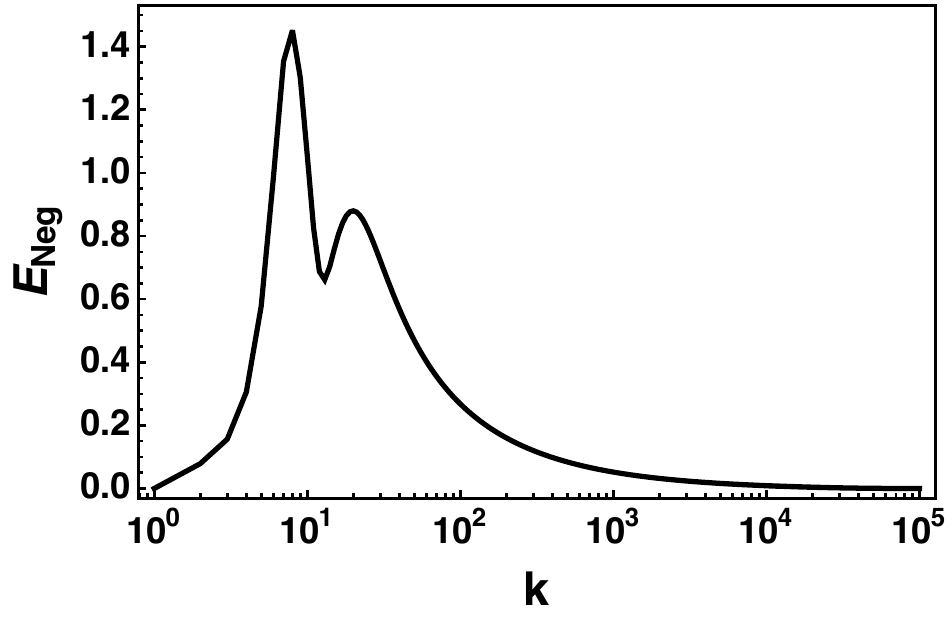} &
{\hspace{0.1in}} &
\includegraphics[width=2.5in,height=1.75in]{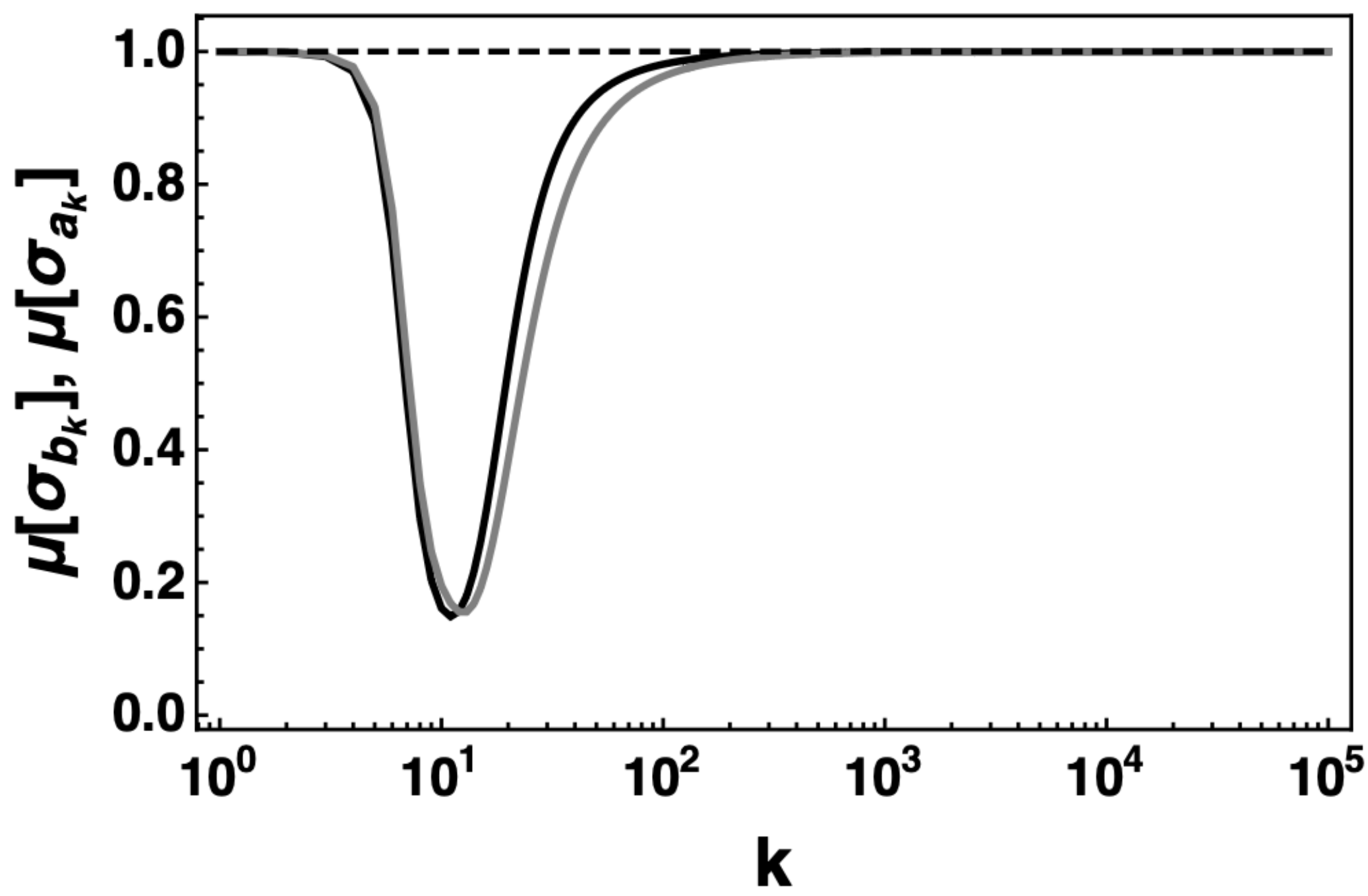}
\end{tabular}
\end{center}
\caption{(left) Log negativity $E_{\mathcal{N}} (\rho_G)$ and (right) purity $\mu_G$
for the full model of BS and single mode squeezers, \Eq{BSSQ:V:eqns:a} and \Eq{BSSQ:V:eqns:b}.}
\label{fig:log:Neg:purity:r:2p5:BSSQ}
\end{figure}

Here we consider the reflectivity profile $R_k = (\Nbk-R_{\textrm{\scriptsize offset}})/N_{b_0}$ as in the previous section.
\Fig{fig:log:Neg:purity:r:2p5:BSSQ} shows the log negativity $E_{\mathcal{N}} (\rho_G)$ 
and the purity $\mu_G$, which show the growth and decay of quantum correlations between the BH and the emitted HR over the course of the evolution.
In this model, the BH begins in a pure ($\mu_G=1$) single mode squeezed state
$\ket{\zeta}_{b_0} = (\cosh r)^{-1/2}\,\sum_{m=0}^{\infty} \left(\tanh(r)/2 \right)^m\,\sqrt{(2 m)!}/m!\ket{2 m}$, which over the course of evolution, thermalizes (de-squeezes) with decreasing occupation number 
($\mu_G<1$), until it completely evaporates to the (pure) vacuum state (again, $\mu_G=1$). Correspondingly, the entanglement, as characterized here by the log negativity $E_{\mathcal{N}} (\rho_G)$, starts at a zero value, becomes non-zero as the BH evolves, and returns to a zero value in its terminal vacuum states.

\begin{figure}[htb]
\centering
\hspace{1.0in}
\includegraphics[width=4.5in,height=2.25in]{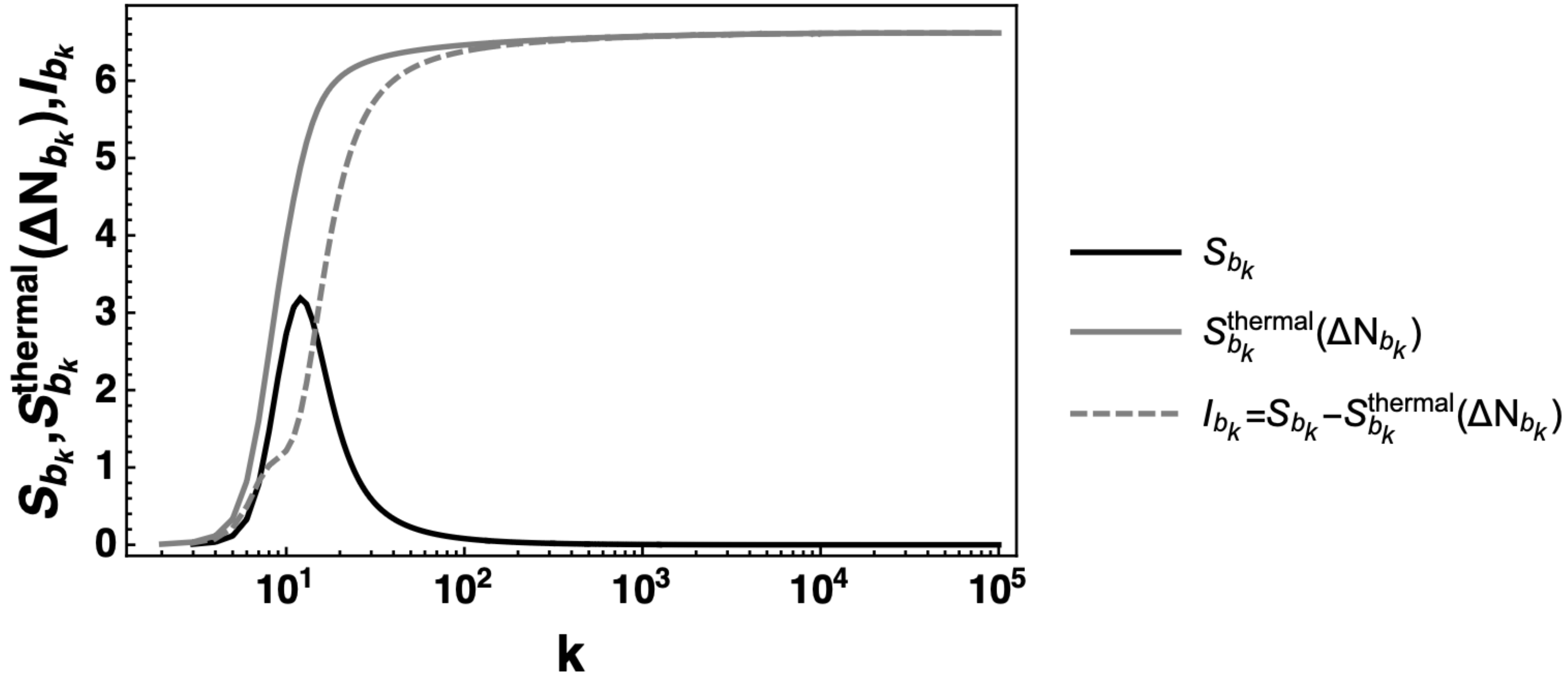}
\caption{Entropy (solid black), effective thermal state entropy (solid gray) and Page Information (dotted gray) for the full model of BS and single mode squeezers, \Eq{BSSQ:V:eqns:a} and \Eq{BSSQ:V:eqns:b}.
}
\label{fig:S:Sthermal:PageInfo:r:2p5:BSSQ}
\end{figure}

In \Fig{fig:S:Sthermal:PageInfo:r:2p5:BSSQ} we address the question of the \textit{Page Information} which conjectures that as the BH evaporates, the information begins to pour out of the BH around the time it is half evaporated. At the same time, the emitted HR is no longer purely thermal, and exhibits correlations. 
Here, the entropy is computed according to \Eq{entropy}. 
The entropy of the effective thermal state $\rho^{thermal}_k$
associated with $\rho_G$ is computed as
\be{eff:thermal:state}
\rho^{thermal}_k \equiv \frac{1}{\Delta  N_{b_k}+1} \, \sum_{m=0}^{\infty}
\left(
\frac{\Delta  N_{b_k}}{\Delta  N_{b_k}+1}
\right)^m
\ket{m}\bra{m}, 
\ee
where $\Delta  N_{b_k} = N_{b_0}-N_{b_k}$.
The Page Information is then defined as
\be{Page:Info}
I_{b_k} = S_V(\rho^{thermal}_{b_k}) - S_V(\rho_{b_k}).
\ee
\Fig{fig:S:Sthermal:PageInfo:r:2p5:BSSQ} shows the characteristics of a \textit{Page Curve} where the Page Information $I_{b_k}$ begins to rapidly rise just as the entropy nears the $S_V(\rho_{b_k})$ maximum, and similarly the entanglement, as here characterized by the log negativity  $E_{\mathcal{N}} (\rho_G)$ \Eq{log:Neg}
and \Fig{fig:log:Neg:purity:r:2p5:BSSQ}, peaks.

\begin{figure}[htb]
\centering
\includegraphics[width=6.0in,height=2.25in]{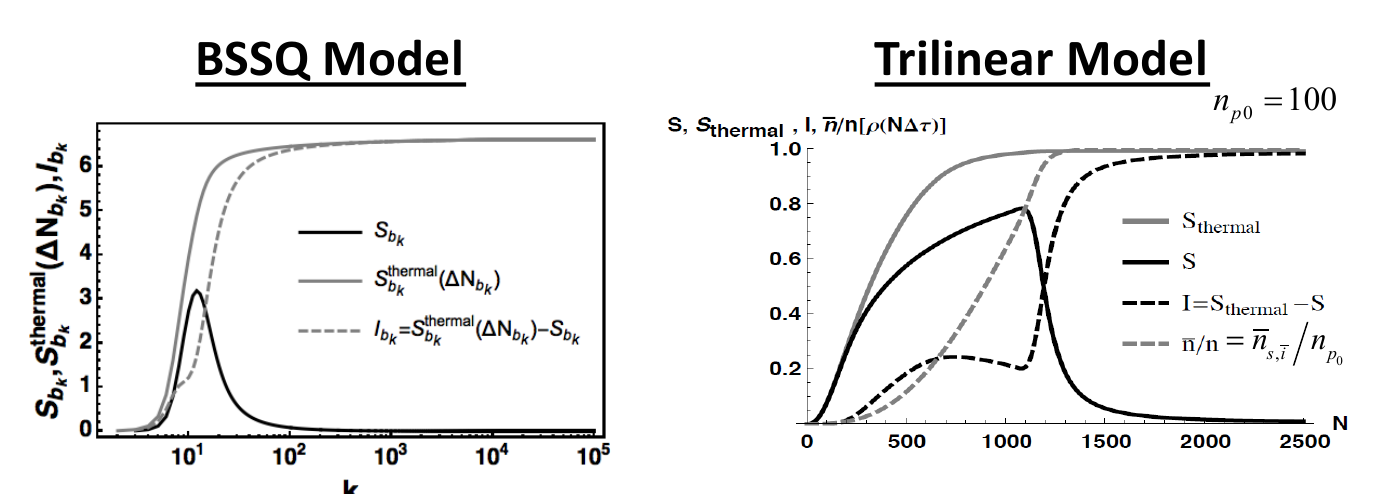}
\caption{Entropy, entropy of effective thermal state and Page information for the for the (left) BSSQ (this work), and (right) Trilinear model \cite{Alsing_CQG:2015}.
}
\label{fig:Page:Info:BSSQ:Trilinear}
\end{figure}

The use $\Delta  N_{b_k} = N_{b_0}-N_{b_k}$ in the definition of the 
effective thermal state $\rho^{thermal}_k$ is motivated by the unitary Trilinear (TL)  Hamiltonian, SPDC (spontaneous parametric down conversion)  with depleted pump model which postulates the fundamental BH/HR interaction as given by \cite{Alsing_CQG:2015,Alsing_Fanto_CQG:2016}
\be{trilinear:Hamiltonian}
H_{p,s,\bar{i}} = r (a_p\,a^\dagger_s\,a^\dagger_{\bar{i}} + a^\dagger_p\,a_s\,a_{\bar{i}}),
\ee
where $p$ the depleting pump models the BH, $s$ the signal models the outgoing HR  
and the idler $\bar{i}$ models the HR partner particle (HRPP) which falls into the BH and acts to decrease the BH mass. Here, $H_{p,s,\bar{i}}$ models the production of HR/HRPP pairs just outside the apparent horizon of the BH, at the cost of occupation number depletion of the BH.
The composite pure state that describes this state is given by
\be{trilinear:state}
\ket{\psi_{TL}} = \sum_{n=0}^\infty c_n\,\ket{n_{p_0}-n}_p\,\ket{n}_s\,\ket{n}_{\bar{i}},
\ee
where $n_{p_0}$ is the large initial occupation number for the BH (taken for simplicity as a Fock state, but also easily modeled as the mean occupation number of a coherent state).
The effective thermal state for this model is given by
$\rho^{thermal}(\bar{n}_p)=(\bar{n}_p~+~1)^{-1}\sum_{m=0}^\infty (\bar{n}_p/(\bar{n}_p+1))^m \ket{m}_p\bra{m}$.
The mean number of "pump" photons $\bar{n}_p$ in this state is given by
\be{np:trilinear}
\bar{n}_p = \bra{\psi_{TL}} a^\dagger_p a_p \ket{\psi_{TL}} =
 n_{p_0} - n \rightarrow N_{b_0} - N_{b_k} = \Delta N_{b_k},
\ee
and therefore one can take
\be{thermal:trilinear:BSSQ}
\rho^{thermal}(\bar{n}_p)\rightarrow\rho_{b_k}(\Delta N_{b_k}).
\ee
The BSSQ model discussed here which uses $\rho_{b_k}(\Delta N_{b_k})$ avoids the detailed modeling of how the HRPP (idler, $\bar{i}$)  is absorbed by the BH (pump, $p$) (but was addressed in \cite{Alsing_CQG:2025}). Note, that at the \textit{cross over point} where $\bar{n}_p\approx\bar{n}_s=n_{\bar{i}}$ and the Page Information begins to be non-zero, the BH state in the trilinear model (when starting in a coherent state of mean occupation number $\bar{n}_p$) is approximately a single mode squeezed state~\cite{Alsing_CQG:2015}. 
A comparison of the entropy, entropy for the effective thermal state and the Page Information for both the BSSQ and Trilinear Model is shown in \Fig{fig:Page:Info:BSSQ:Trilinear}, which exhibit qualitatively similar behavior.
Thus, the association in \Eq{np:trilinear}  and \Eq{thermal:trilinear:BSSQ} is well motivated.

\subsubsection{Temporal AutoCorrelation Plots}\label{sec:ent:plots}
In this section we investigate the temporal correlations between early and late 
HR $\langle X^{\pm}_{a^{out}_{k}}\,X^{\pm}_{a^{out}_{k-\Delta k}} \rangle$,
and the state of the 
BH $\langle X^{\pm}_{b_{k}}\,X^{\pm}_{b_{k-\Delta k}} \rangle$.
\begin{figure}[ht]
\begin{tabular}{cc}
\includegraphics[width=3.0in,height=1.75in]{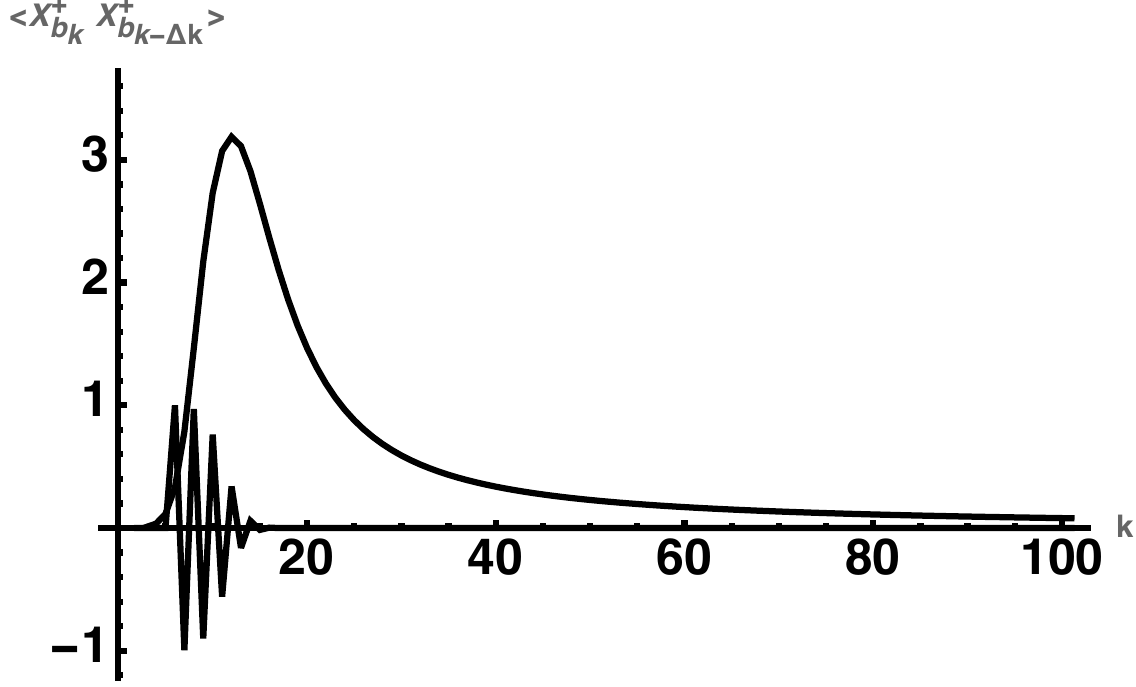} &
\includegraphics[width=3.0in,height=1.75in]{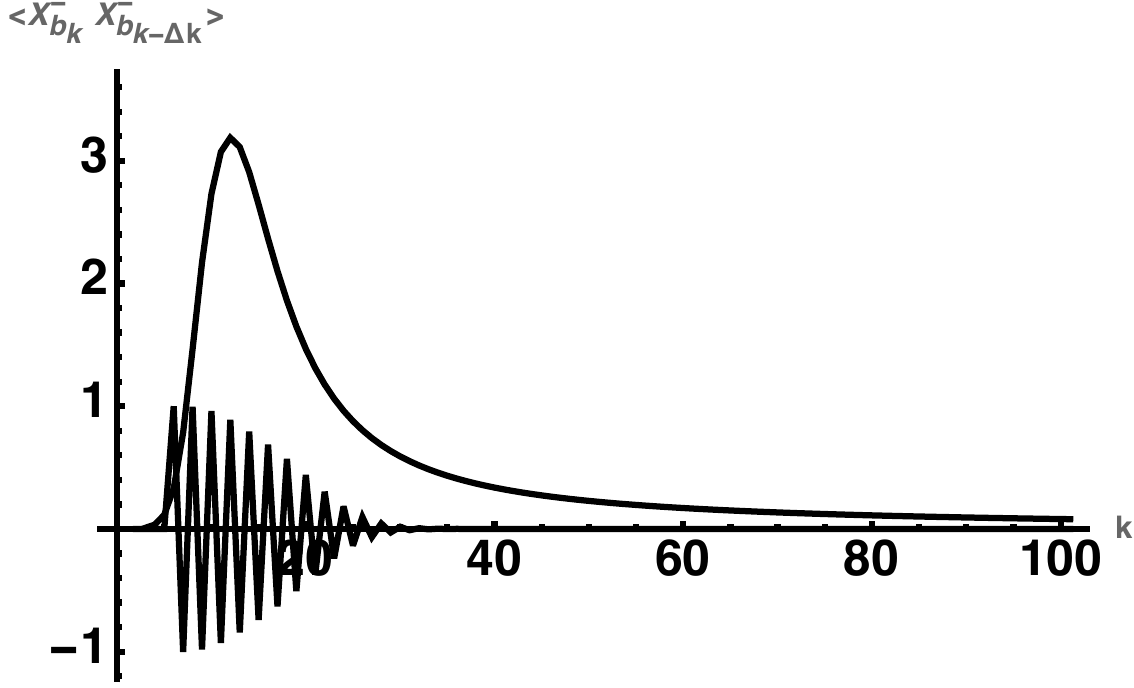}\\
\includegraphics[width=3.0in,height=1.75in]{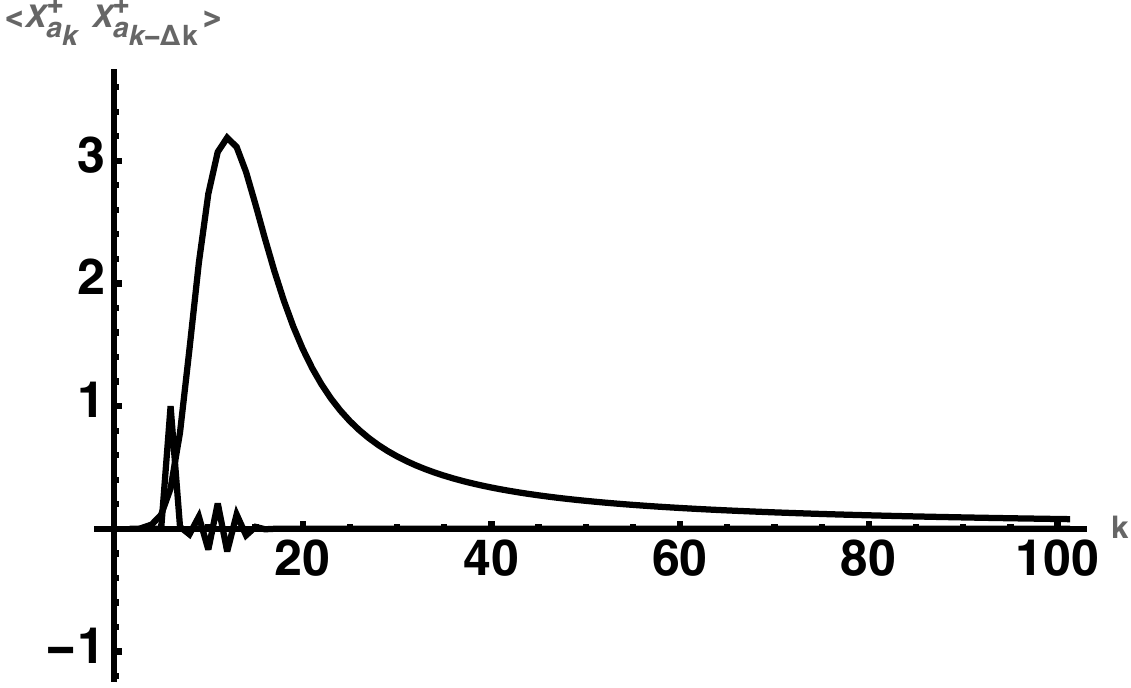} &
\includegraphics[width=3.0in,height=1.75in]{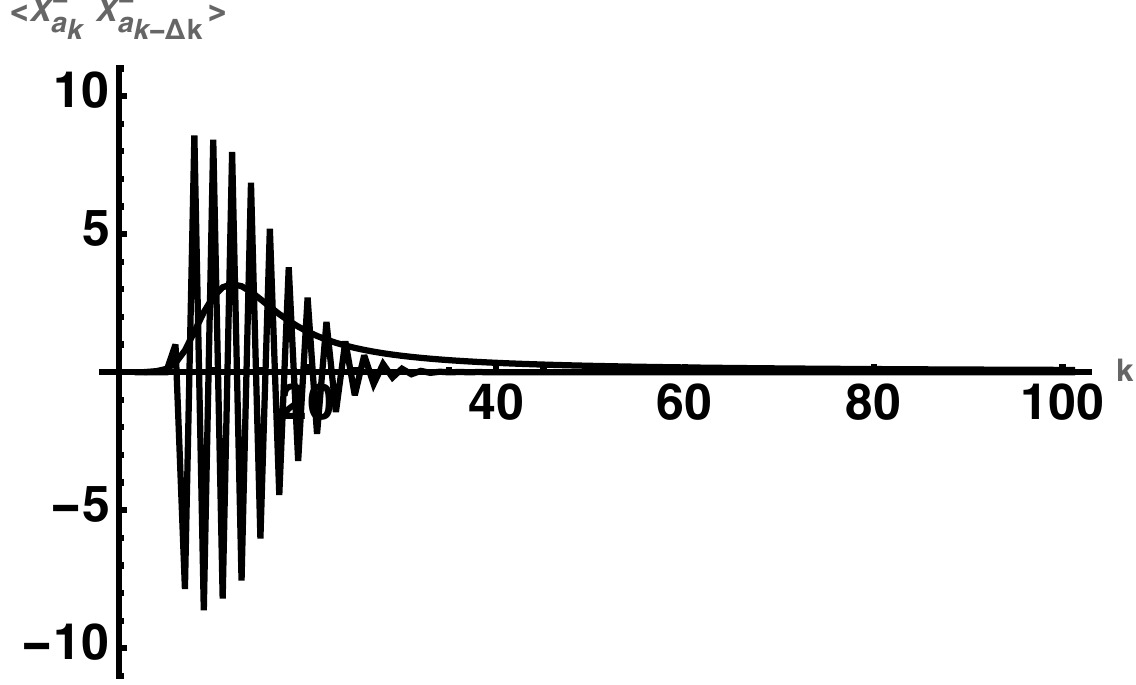}
\end{tabular}
\caption{(top row) Autocorrelation functions $\langle X^{\pm}_{b_{k}}\,X^{\pm}_{b_{k-\Delta k}} \rangle$
and (bottom row) $\langle X^{\pm}_{a_{k}}\,X^{\pm}_{a_{k-\Delta k}} \rangle$ for $r=2.5$.
plotted against the BH entropy $S_{b_k}$.}
\label{fig:AutoCorrelations:r:2p5:kstart:5}
\end{figure}
These autocorrelation functions are computed by recursively expanding the quadrature operators $\Xakoutpm$ and $\Xbkpm$
in \Eq{BSSQ:X:eqns:a} and \Eq{BSSQ:X:eqns:b}, respectively, in terms of initial quadratures $\Xakinpm$ and $X^{\pm}_{b_{0}}$.
The useful formulas are detailed but not illuminating, and their derivation is relegated to the Appendix A.
Note that each of these initial quadratures are only correlated with themselves (i.e. all cross correlations are zero) and that 
$\langle (X^{\pm}_{a^{in}_{k}})^2\rangle = V^{\pm}_{a^{in}_{k}} = 1$,
and
$\langle (X^{\pm}_{b_{0}})^2\rangle = V^{\pm}_{b_{0}} = e^{\mp 2 r}$.
In \Fig{fig:AutoCorrelations:r:2p5:kstart:5} we plot these auto correlations functions on top of the evolution of the entropy $S_{b_k}$ of the BH.
We see that anti-squeezed correlations $\langle X^{-}_{i_{k}}\,X^{-}_{i_{k-\Delta k}} \rangle$ for $i\in \{a,b\}$ are stronger and more persistent than the shorted lived squeezed correlations $\langle X^{+}_{i_{k}}\,X^{+}_{i_{k-\Delta k}} \rangle$. 

\begin{figure}[ht]
\begin{tabular}{cc}
\includegraphics[width=3.0in,height=1.75in]{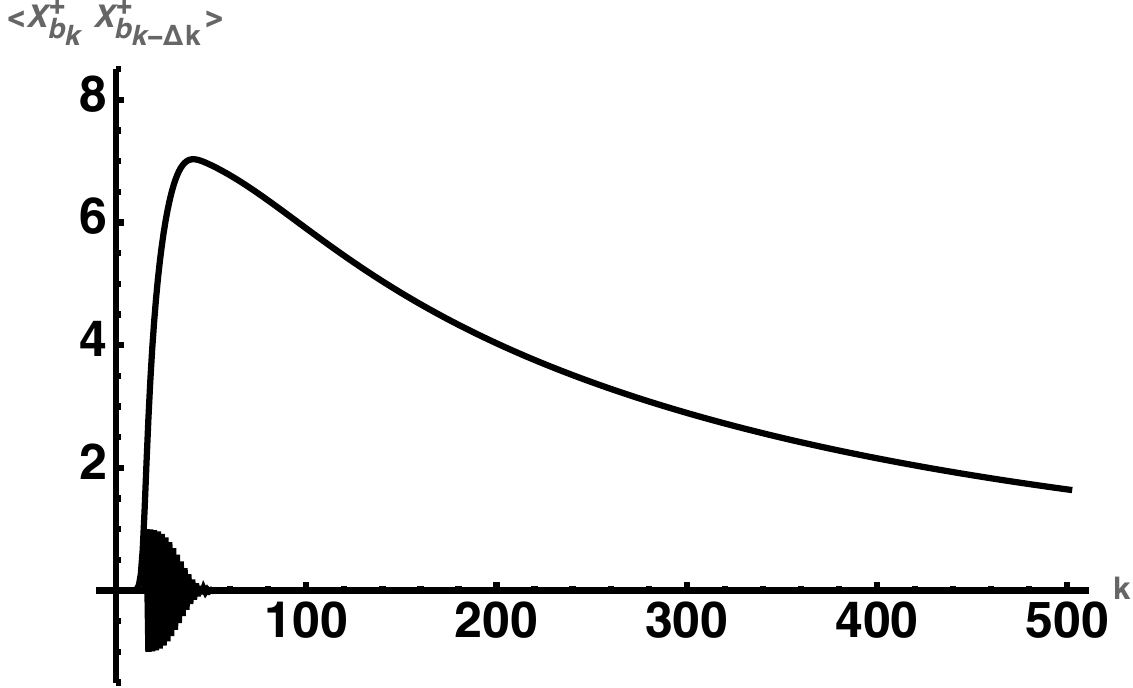} &
\includegraphics[width=3.0in,height=1.75in]{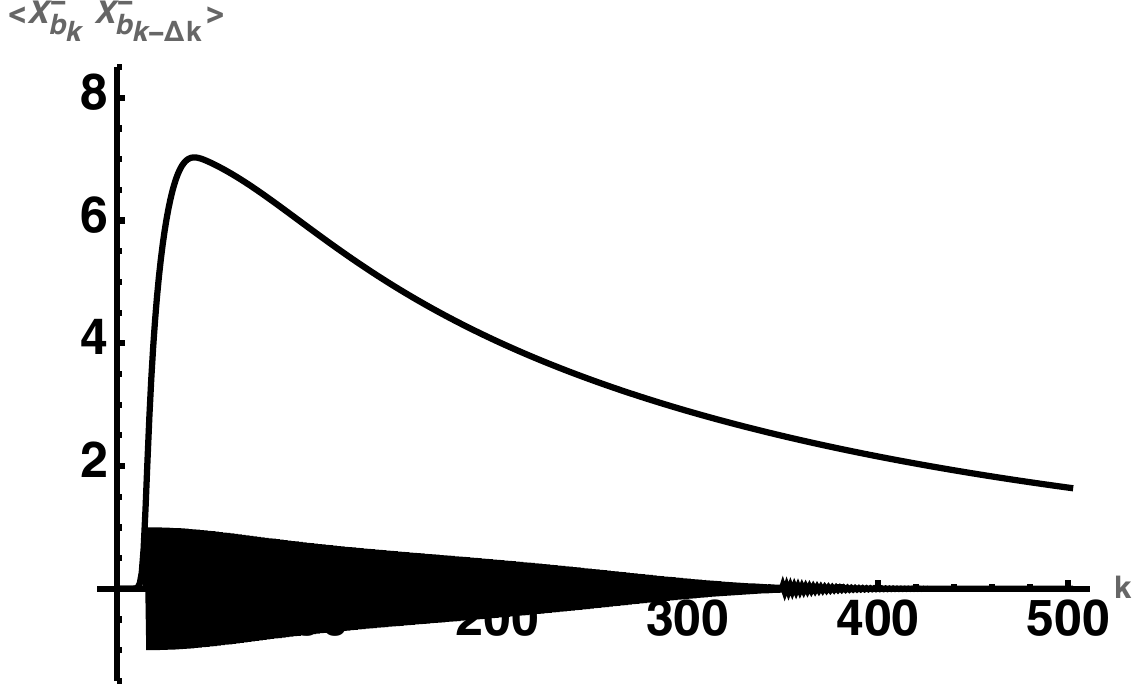}\\
\includegraphics[width=3.0in,height=1.75in]{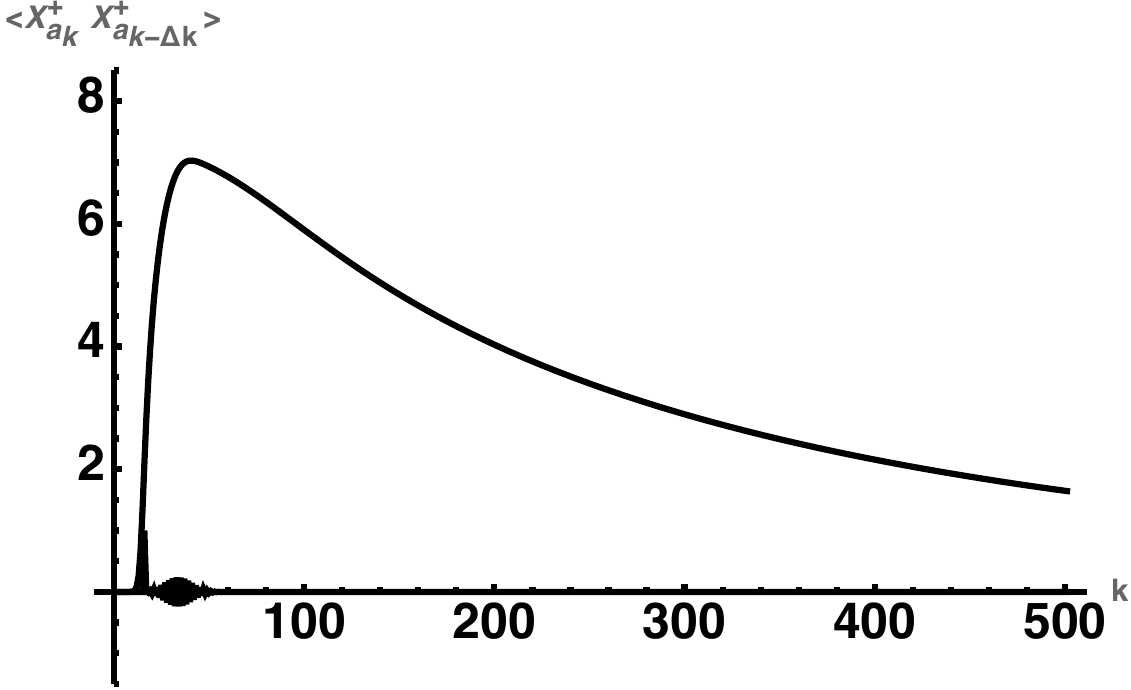} &
\includegraphics[width=3.0in,height=1.75in]{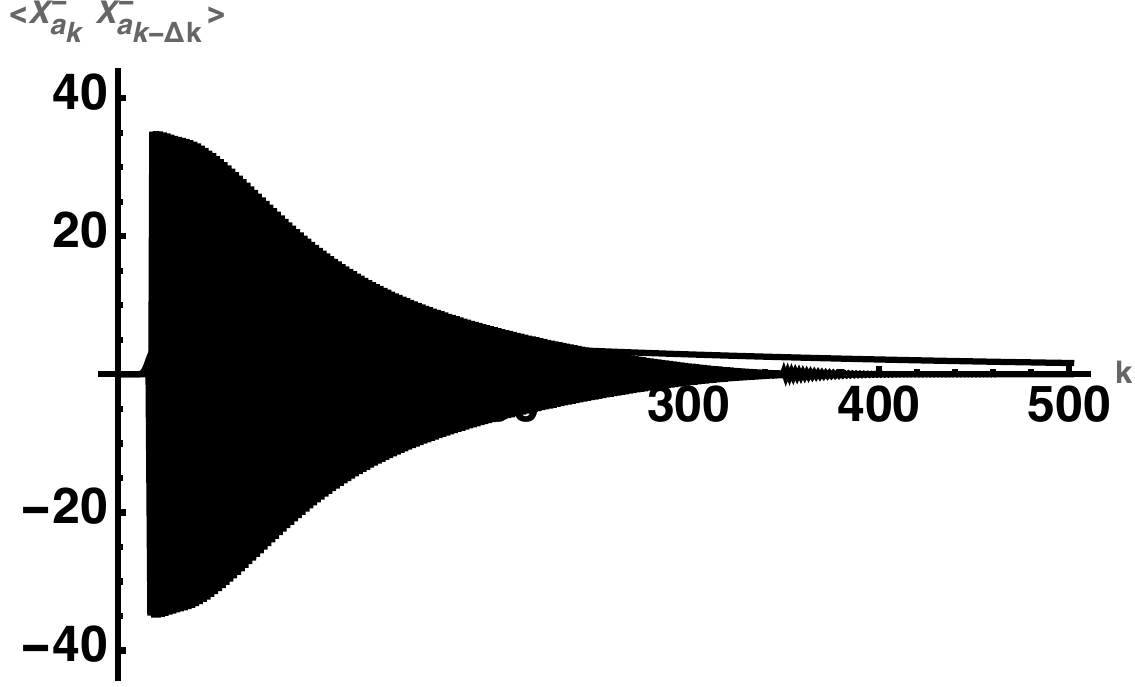}
\end{tabular}
\caption{(top row) Autocorrelation functions $\langle X^{\pm}_{b_{k}}\,X^{\pm}_{b_{k-\Delta k}} \rangle$
and (bottom row) $\langle X^{\pm}_{a_{k}}\,X^{\pm}_{a_{k-\Delta k}} \rangle$ for $r=5.0$.
plotted against the BH entropy $S_{b_k}$. The oscillations are so fast that they appear in the figures as a solid black interiors, although the more slowly varying envelopes are clearly evident.}
\label{fig:AutoCorrelations:r_5:kstart:15}
\end{figure}
The same behavior is exhibited for the case of a more strongly initially squeezed BH, given by $r=5.0$ and shown in 
\Fig{fig:AutoCorrelations:r_5:kstart:15}, for which the BH entropy $S_{b_k}$ evolves over longer iterations.
The $\langle X^{-}_{b_{k}}\,X^{-}_{b_{k-\Delta k}} \rangle$ shows that the BH exhibits correlations between its early and late evolution. Similarly, $\langle X^{-}_{a^{out}_{k}}\,X^{-}_{a^{out}_{k-\Delta k}} \rangle$ also shows correlations between early and late time HR, due to the entangling of the HR and BH states at iteration $k-\Delta k$ influencing the state of the BH, and hence influencing the subsequent entanglement of the HR and BH at the later iteration $k$. The information in the BH emerges in these correlations, as suggested by Page.

\section{Conclusion and Future Directions}\label{sec:conclusion:future:dirs}
The above investigations were motivated by developing quantum optical inspired models for unitary BH evaporation models that could reproduce the salient feature of the Page Curve, as well as have the BH completely evaporate. As discussed above the free parameters in the BSSQ model presented here are the set of BS reflectivities $\{R_k\}$. Success at reproducing the Page Curve came from choosing these reflectivities such that in early stages of evolution the BH was nearly opaque ($R_k\approx 1$) while at later stages the BH was nearly transparent ($R_k\approx 0$). A more satisfactory model would incorporate some additional physical principle (e.g. current thoughts on BH emission rates for different classes of BH evaporation models) that would have the reflectivities determined self consistently (i.e. possibly coupled to the evolution of the variances $\Vakpm$ and $\Vbkpm$, or to an external BH evolution equation). 

For example, one could argue as follows. Let 
$E_{a_1} = \int_0^\infty d\omega\, \hbar\, \omega N_{a_1} = \int_0^\infty d\omega\, \hbar\, \omega/(e^{\beta\hbar\omega}-1)$ 
be the total energy emitted into HR frequencies after iteration $k=1$, where 
$\beta = 1/(k_B T_{BH}) = 2\pi c/(\hbar \kappa) = 8 \pi G M_{BH}/(\hbar c^3)$ is the inverse temperature of the BH. 
A simple integration yields $E_{a_1}  = \hbar \pi^2 /(6 (\hbar\beta)^2)\sim 1/M^2_{BH}$. 
On the other hand, we also have that after one iteration $E_{b_1}\approx R_1\,\sinh^2(r) = R_1 N_{b_0} \sim R_1 M_{BH}$. 
%
Using $E_{b_1} = E_{b_0}-E_{a_1}\Rightarrow R_1\,M_{BH} \sim M_{BH} - c_1/M^2_{BH}$ for some constant $c_1$ gives us $R_1\sim 1- c_1/M^3_{BH}$,  valid at least for early times.
Since $E_{b_1} \sim M_{BH} - c_1\,M^{-2}_{BH}$ 
we have that the heat capacity, given by $dE_{b_1}/dT_{BH} = -M^2_{BH}\,dE_{b_1}/dM_{BH} = -M^2_{BH}\,(M_{BH} + 2\,c_1\,M^{-3}_{BH}) < 0$, is negative. In fact, one would obtain a negative heat capacity for any $E_{a_1}\sim M^{-p}_{BH}$ with $p>1$.
However, numerical simulations (not presented here) reveal that the BH evaporation exhibits the same bursting phenomena exhibit in the case of $R_k=R=$constant for all $k$ examined previously, only this time much more rapidly and intensely. Much of this is due to choosing $R_k\sim 1- M^{-p}_{BH}$ with $p>1$ and large initial value of $r$ so that $R_k\sim1$ for early times, but transitions very rapidly to near zero at some midpoint of the evolution, 
i.e. the BH is essentially transparent initially, and subsequently the radiation "bursts out of the cavity," rather than slowly leaking out initially. For this reason, we decided to explore both the case $R_k = 1-\ep$ for all $k$, as well as 
$R_k = (\Nbk-R_{\trm{\scriptsize offset}})/N_{b_0}$

In addition, the freedom to choose the reflectivities $\{R_k\}$ should also allow one to model other BH evaporation models, for example remnant models where the BH does not completely evaporate, or possibly even models where the BH mass (say in a collapsing shell) never crosses its own (decreasing) Schwarzschild radius. The goal of the current BSSQ model is to capture the essential feature of these and other BH evaporation models in a simple and tractable unitary BH evolution model, while reproducing essential features, include the Page Curve.

\ack The author would like to acknowledge Timothy Ralph and Agata Branczyk 
for useful discussions on the initial conceptualization of quantum optical models for black hole evaporation.

\noindent The author has  no competing interests for this work.
%

\noindent The \tit{Mathematica} codes that support the data used in this paper and other finding are not currently publicly available. The author intends to make data openly available in the near future at 
{\verb+https://dataverse.harvard.edu/dataverse/alsingpm_research+}. 
\appendix
\section{Details Calculations}\label{app}
\subsection{Temporal Autocorrelation functions}\label{app:autocorrelations}
The quadratures for each iteration $k$ are given in \Eq{BSSQ:X:eqns:a} and \Eq{BSSQ:X:eqns:b} and repeated below for convenience
\bea{app:BSSQ:X:eqns}
\Xakoutpm &=& e^{\mp\rak} \left[ \sqrt{\Rk} \, \Xakinpm  + \sqrt{(1-\Rk)}\, \Xbkmpm \right], \label{app:BSSQ:X:eqns:a}\\
\Xbkpm &=& e^{\mp\rbk} \left[  \sqrt{(1-\Rk)} \, \Xakinpm  -      \sqrt{\Rk}\, \Xbkmpm \right]. \label{app:BSSQ:X:eqns:b}
\eea
A straightforward, but tedious, iterative expansion leads to the expressions for the quadratures in terms of the incoming vacuum modes $\Xakinpm$ and the initial BH variances~$X^{\pm}_{b_{0}}$
\be{app:Xbpm}
\hspace{-1.0in}
\Xbkpm = e^{\mp r_{b_{k}}}\,\sqrt{1-R_{k}}\,X^{\pm}_{a^{in}_{k}} 
+ \sum_{j=1}^{k-1} \mathcal{F}^{(\pm)(k)}_{j}\,X^{\pm}_{a^{in}_{k-j}} + \mathcal{G}^{(\pm)(k)}\,X^{\pm}_{b_{0}},
\ee
and
\bea{app:Xapm}
\hspace{-0.25in}
\Xakoutpm &=& 
e^{\mp r_{a_{k}}}\,\sqrt{R_{k}}\,X^{\pm}_{a^{in}_{k}} + e^{\mp r_{a_{k}}}\,\sqrt{1-R_{k}}
\Big( 
e^{\mp r_{b_{k-1}}}\,\sqrt{1-R_{k-1}}\,X^{\pm}_{a^{in}_{k-1}} \qquad
 \no
{} & & 
\hspace*{2.25in}
+ \sum_{j=1}^{k-2} \mathcal{F}^{(\pm)(k-1)}_{j}\,X^{\pm}_{a^{in}_{k-1-j}} + \mathcal{G}^{\pm(k-1)}\,X^{\pm}_{b_{0}}
\Big),\qquad\quad
\eea
where we have defined the functions
\bea{app:F:G}
\mathcal{F}^{(\pm)(k)}_{j} &=& (-1)^{j}\,\left( e^{\mp r_{b_{k-j}}} \,\sqrt{1-R_{k-j}}\right)\,
\prod_{\ell=1}^{j} \left( e^{\mp r_{b_{k+1-\ell}}}\,\sqrt{R_{k+1-\ell}}\right), \\
\mathcal{G}^{(\pm)(k)} &=& (-1)^{k}\,\prod_{j=1}^{k-1} \left( e^{\mp r_{b_{k-j}}}\,\sqrt{R_{k-j}}\right).
\eea
Note that incoming vacuum modes $X^{\pm}_{a^{in}_{k}}$ are all independent of each other, 
and the initial BH quadratures $X^{\pm}_{b_{0}}$,  with variances 
$\langle(X^{\pm}_{a^{in}_{k}})^2\rangle= \Vakinpm =1$.
The initial BH variances are given by $\langle (X^{\pm}_{b_{0}})^2\rangle = V^{\pm}_{b_{0}} = e^{\mp 2 r}$, 
with $r\equiv r_{b_0}$ the squeezing parameter of the initial BH state.

We are now interested in forming the early/late time HR and BH (auto)correlations 
 $\langle X^{\pm}_{a^{out}_{k}}\,X^{\pm}_{a^{out}_{k-\Delta k}} \rangle$,
 $\langle X^{\pm}_{b_{k}}\,X^{\pm}_{b_{k-\Delta k}} \rangle$, respectively.
 For a fixed $\Delta k$ we can write the quadratures in these correlation functions  for $k$ and $k-\Delta k$ in terms of the incoming vacuum modes $\Xakinpm$ and the initial BH variances $X^{\pm}_{b_{0}}$, by using  \Eq{app:Xbpm} and \Eq{app:Xapm}. Only the overlapping identical vacuum and initial BH quadratures lead to non-zero variances
  $\langle (X^{\pm}_{a^{in}_{j}})^2\rangle = V^{\pm}_{a^{in}_{j}}=1$, and 
 $\langle (X^{\pm}_{b_{0}})^2\rangle = V^{\pm}_{b_{0}} = e^{\mp 2 r}$.
 By breaking up sum over $j$ that appear in  \Eq{app:Xbpm} and \Eq{app:Xapm} into
 $j=1$ to $j=\Delta k-1$, and $j=\Delta k$ to its upper limit, we arrive (after some lengthy but straightforward algebra) at the expressions
\be{XbkXbkmdk}
\hspace{-1.5in}
\langle X^{\pm}_{b_{k}}\,X^{\pm}_{b_{k-\Delta k}} \rangle =
\mathcal{F}^{(\pm)(k)}_{\Delta k}\,\left( e^{\mp r_{b_{k-\Delta k}}}\,\sqrt{1-R_{k-\Delta k}}  \right)
+ \sum_{j=1}^{k-\Delta k-1} \mathcal{F}^{(\pm)(k)}_{j+\Delta k}\,\mathcal{F}^{(\pm)(k-\Delta k)}_{j}
+ \mathcal{G}^{(\pm)(k)}\,\mathcal{G}^{(\pm)(k-\Delta k)}\,e^{\mp 2 r},\;\quad
\ee
and
\bea{XakXakmdk}
\hspace{-1.0in}
\langle X^{\pm}_{a^{out}_{k}}\,X^{\pm}_{a^{out}_{k-\Delta k}} \rangle &=& \left( e^{\mp 2 r_{a_{k}}}\,\sqrt{1-R_{k}}  \right)\,
\Big[ \mathcal{F}^{(\pm)(k-1)}_{\Delta k-1}\,\left( e^{\mp 2 r_{a_{k-\Delta k}}}\,\sqrt{R_{k-\Delta k}} \right) \no
\hspace{-1.0in}
&+& \mathcal{F}^{(\pm)(k-1)}_{\Delta k}\,\left( e^{\mp 2 r_{a_{k-\Delta k}}}\,\sqrt{1-R_{k-\Delta k}} \right) \,
\left( e^{\mp 2 r_{b_{k-\Delta k-1}}}\,\sqrt{1-R_{k-\Delta k-1}} \right), \no
\hspace{-0.75in}
&+&  \sum_{j=1}^{k-\Delta k-2} \mathcal{F}^{(\pm)(k-1)}_{j+\Delta k}\,\mathcal{F}^{(\pm)(k-\Delta k-1)}_{j}
+ \mathcal{G}^{(\pm)(k-1)}\,\mathcal{G}^{(\pm)(k-\Delta k-1)}\,e^{\mp 2 r} \Big].
\eea
\Eq{XbkXbkmdk} and \Eq{XakXakmdk} are the formulas used to plot the autocorrelations in \Sec{sec:ent:plots}.
Cross HR/BH such as 
$\langle X^{\pm}_{a^{out}_{k}}\,X^{\pm}_{b_{k-\Delta k}} \rangle$ and 
$\langle X^{\pm}_{b_{k}}\,X^{\pm}_{a^{out}_{k-\Delta k}}\, \rangle$
can be also computed in a similar fashion (but are not shown or plotted here).
%
\section*{References}

\begin{thebibliography}{10}
\expandafter\ifx\csname url\endcsname\relax
  \def\url#1{{\tt #1}}\fi
\expandafter\ifx\csname urlprefix\endcsname\relax\def\urlprefix{URL }\fi
\providecommand{\eprint}[2][]{\url{#2}}

\bibitem{Hawking:1975}
Hawking S~W 1975 {\em Commun. Math. Phys.\/} {\bf 43} 199

\bibitem{Page:2005}
Page D~N 2005 {\em New J. Phys.\/} {\bf 7} 203

\bibitem{Almheiri_Hartman:2020}
Almheiri A, Hartman T, Maldacena J, Shaghoulian E and Tajdini A 2020 {\em J.
  High Energy Phys.\/} {\bf 05} 013 (arxiv:1911.12333)

\bibitem{Almheiri_Hartman:2021}
Almheiri A, Hartman T, Maldacena J, Shaghoulian E and Tajdini A 2021 {\em Rev.
  Mod. Phys.\/} {\bf 93}(12) 035002 (arxiv:2006.06872v1)

\bibitem{Page:1993a}
Page D~N 1993 {\em Phys.Rev.Lett.\/} {\bf 71} 1291 (arxiv:gr--qc:9305007v2)

\bibitem{Page:1993b}
Page D~N 1993 {\em Phys.Rev.Lett.\/} {\bf 71} 3743 (arxiv:gr--qc:9306083v2)

\bibitem{Harlow:2016}
Harlow D 2016 {\em Rev. Mod. Phys.\/} {\bf 88} 015002 (arxiv:14091231v5)

\bibitem{Ydri:2025}
Ydri B 2025 {\em Lectures on General Relativity, Cosmology and Quantum Black
  Holes\/} (IOP Books)

\bibitem{Cox_Forshaw:2022}
Cox B and Forshaw J 2022 {\em Black Holes\/} (Harper Collins, NY)

\bibitem{Alsing_CQG:2025}
Alsing P~M 2025 {\em Classical and and Quantum Gravity\/} {\bf 42}(7) 075019

\bibitem{Ryu_Takayanagi:2006}
Ryu S and Takayanagi T 2006 {\em J. High Energy Phys.\/} {\bf 08} 45
  (hep--th/0605073)

\bibitem{HRT:2007}
Hubney V~E, Rangamani M and Takayanagi T 2007 {\em J. High Energy Phys.\/} {\bf
  0707} 62 (arxiv:0705.0016 [hep--th])

\bibitem{Maldacena_Susskind:2013}
Maldacena J and Susskind L 2013 {\em Fortsch. Phys.\/} {\bf 61} 781

\bibitem{Almheiri:2020}
Almheiri A, Engelhardt N, Marolf D and Maxfield H 2019 {\em J. High Energy
  Phys.\/} {\bf 2019} 063(arxiv:1905:08762)

\bibitem{Penington:2020}
Penington G 2020 {\em J. High Energy Phys.\/} {\bf 2020}(9) (arxiv:1905:08255)

\bibitem{Penington:2022}
Penington G, Shenker S~H, Stanford D and Yang Z 2022 {\em J. High Energy
  Phys.\/} {\bf 2022}(205) (arxiv:1911.11977 [hep--th])

\bibitem{Nation_Blencowe:2010}
Nation P~D and Blencowe M~P 2010 {\em New J. Phys.\/} {\bf 12} 095013

\bibitem{Adami_VerSteeg:2014}
Adami C and Steeg G~V 2014 {\em Class. Quantum Grav.\/} {\bf 31} 075015
  (arxiv:gr--qc/0407090v8)

\bibitem{Bradler_Adami:2016}
Br\'{a}dler K and Adami C 2016 {\em Phys. Rev. Lett.\/} {\bf 116} 101301
  (arXiv:1505.0284)

\bibitem{Alsing_CQG:2015}
Alsing P~M 2015 {\em Classical and and Quantum Gravity\/} {\bf 32}(7) 075010

\bibitem{Alsing_Fanto_CQG:2016}
Alsing P~M and Fanto M~L 2016 {\em Classical and and Quantum Gravity\/} {\bf
  33}(1) 015005

\bibitem{Scully_Zubairy:1997}
Scully M~O and Zubairy M~S 1997 {\em Quantum Optics, (Chap. 9)\/} (Cambridge:
  Cambridge University Press)

\bibitem{Agarwal:2013}
Agarwal G~S 2013 {\em Quantum Optics\/} (Cambridge University Press, Cambridge)

\bibitem{Gerry_Knight:2023}
Gerry C~C and Knight P~L 2023 {\em Introductory Quantum Optics, 2nd Ed.\/}
  (Cambridge University Press, Cambridge)

\bibitem{Rice:2025}
Rice P 2025 {\em An Introduction to Quantum Optics: an Open Systems Approach,
  2nd ed.\/} (IOP Books)

\bibitem{Adesso:Illuminati:2007}
Adesso G and Illuminati F 2007 {\em Journal of Physics A: Mathematical and
  Theoretical\/} {\bf 40} 7821--7880

\bibitem{Weedbrook_GS_RMP:2012}
Weedbrook C, Pirandola S, RGarcia-Patron, Cerf N, Ralph T, Shapiro J and Lloyd
  S 2012 {\em Rev. Mod. Phys.\/} {\bf 84} 621

\bibitem{Tserkis_Ralph:2017}
Tserkis S and Ralph T 2017 {\em Phys. Rev. A\/} {\bf 96} 062338
  (arxiv:1705.03612)

\bibitem{Serafini:book:2017}
Serafini A 2017 {\em Quantum Continuous Variables: A primer of theoretical
  methods\/} (Taylor \& Francis)

\end{thebibliography}
\providecommand{\noopsort}[1]{}\providecommand{\singleletter}[1]{#1}%
\providecommand{\newblock}{}

\end{document}